\PassOptionsToPackage{sort&compress}{natbib}
\documentclass[review,12pt,3p]{elsarticle}

\makeatletter
\def\ps@pprintTitle{%
	\let\@oddhead\@empty
	\let\@evenhead\@empty
	\def\@oddfoot{}%
	\let\@evenfoot\@oddfoot}

\usepackage{multirow}
\usepackage[
singlelinecheck=false 
]{caption}
\usepackage{enumitem}

\usepackage{verbatim}
\usepackage{enumitem}
\usepackage{bm}
\usepackage{setspace}
\usepackage{makecell}
\usepackage{booktabs} 
\usepackage{numprint} 

\usepackage{amsmath}

\usepackage{amssymb}
\usepackage{color}
\usepackage{xcolor}

\begin{document}
    
    \renewcommand{\arraystretch}{1.3}
		
    \begin{frontmatter}
    
    \title{Percolation of random compact diamond-shaped systems on the square lattice}
    
    \author[label2]{Charles S. do Amaral \corref{cor1}}
    \address[label2]{Departamento de Matem\'atica - Centro Federal de Educação Tecnológica de Minas Gerais, \linebreak Av. Amazonas 7675, Belo Horizonte, MG, Brasil.}
    \ead{charlesmat@cefetmg.br}
    \cortext[cor1]{Corresponding author.}
    \author[label1]{Mateus G. Soares}		
    \address[label1]{Departamento de Computação - Centro Federal de Educação Tecnológica de Minas Gerais, Av. Amazonas 7675, Belo Horizonte, MG, Brasil.}
    \author[label3]{Robert M. Ziff}		
    \address[label3]{Center for the Study of Complex Systems and Department of Chemical Engineering, University of Michigan, Ann Arbor, Michigan 48109-2800, USA}

    \date{\today}
    
    \begin{abstract}  
        We study site percolation on a square lattice with random compact diamond-shaped neighborhoods. Each site $s$ is connected to others within a neighborhood in the shape of a diamond of radius $r_s$, where $r_s$ is uniformly chosen from the set $\{i, i+1, \ldots, m\}$ with $i \leq m$. The model is analyzed for all values of $i = 0, \ldots, 7$ and $m = i, \ldots, 10$, where $\overline{z}(i,m)$ denotes the average number of neighbors per site and $p_c(i,m)$ is the critical percolation threshold. For each fixed $i$, the product $\overline{z}(i,m)\,p_c(i,m)$ is found to converge to a constant as $m \to \infty$. Such behavior is expected when $i=m$ (single diamond sizes), for which the product $\overline{z}(i)\,p_c(i)$ tends toward $2^d\eta_c$, where $\eta_c$ is the continuum percolation threshold for diamond-shaped regions or aligned squares in two dimensions ($d=2$). This case is further examined for $i = 1, \ldots, 10$, and the expected convergence is confirmed. The particular case $i = m$ was first studied numerically by Gouker and Family in 1983.  We also study the relation to systems of deposited diamond-shaped objects on a square lattice.  For monodisperse diamonds of radius $r$, there is a direct mapping to percolation with a diamond-shaped neighborhood of radius $2r+1$, but when there is a distribution of object sizes, there is no such mapping.  We study the case of mixtures of diamonds of radius $r=0$ and $r=1$, and compare it to the continuum percolation of disks of two sizes. 
    \end{abstract}
    
    \end{frontmatter}

    \section{Introduction}

	The \textit{Bernoulli site percolation model} was introduced by Broadbent and Hammersley in 1957 \cite{broadbent} with the idea of modeling the flow of a deterministic fluid through a random porous medium. In this model, we assign probabilities $p$ and $1-p$, independently, so that each site of an infinite and connected graph is \textit{open} and \textit{closed}, respectively. If a site is open, then fluid can pass through it; otherwise, the fluid passage is blocked. This model presents a geometric phase transition characterized by the emergence of an infinite sequence of connected open sites for values of $p$ greater than a certain constant $p_c$, called the \textit{critical point} or \textit{percolation threshold}. Instead of sites, we can assign probabilities to the bonds to be open or closed. In this case, the model is referred to as the \textit{Bernoulli bond percolation model}. 
	
	Variations of this model have subsequently emerged, finding applications in numerous scientific fields \cite{stauffer94,araujo,saberi,hassan, sahimi}. For instance, percolation on lattices with extended neighborhoods has been applied to problems of deposition of extended shapes on a lattice \cite{koza, koza2} and has connections to spatial models for the spread of epidemics via long-range links \cite{ziff, sander}. Furthermore, this lattice structure occupies an intermediate position between discrete percolation and continuum percolation, making further research essential for establishing relationships between these models \cite{koza, koza2, domb_1972, xun, xun2}.
	
	Studies of percolation models with extended neighborhoods generally examine the relationship between the critical point $p_c$ and the coordination number $z$ or other lattice characteristics. It is well-established that, for these models, the value of $p_c$ is linked to the continuum percolation threshold $\eta_c$ \cite{penrose, koza, koza2, domb_1972, iribarne_1999}, particularly when the coordination number is large and the continuum objects have the same shape as the lattice neighborhood. This relationship is expressed as
	\begin{align}
		zp_c \sim 2^d \eta_c
		\label{cont}
	\end{align}
    \noindent in which $d$ is the system dimension. For example, in cases where each site's neighborhood is composed of compact diamond-shaped structures of the same size on a square lattice with $d=2$, Xun et al.\ \cite{xun} plotted the graph of $z$ against ${1}/{p_c}$ using critical values for $p_c$ obtained from Gouker and Family \cite{gouker}, and observed a linear behavior with a slope of approximately $4.175$. Given that $\eta_c \approx 1.09884280(9)$ for aligned squares as reported by Mertens \cite{mertens}, the obtained slope is in relatively close agreement with the prediction in Eq.~(\ref{cont}) ($2^{d} \eta_c \approx 4.3954$). Here, we follow the assumption made in~\cite{jasna2024} that, for the continuum percolation model, a diamond-shaped region on the square lattice is equivalent to a square-shaped region on a diagonally oriented square lattice, and therefore both configurations share the same continuum percolation threshold. 

    To account for finite-$z$ corrections, the authors in Ref.~\cite{xun} also propose the empirical expression $p_c = {c}/{(z + b)}$, where $b$ is a fitting parameter and $c = 2^d \eta_c$. This expression can be rearranged as
    \begin{equation}
    z = \frac{c}{p_c} - b,
    \label{eq_invpc}
    \end{equation}
    which yields a linear relationship between $z$ and $1/p_c$.
    
    Therefore, by plotting $z$ versus $1/p_c$, one can extract an estimate of $2^d\eta_c$ from the slope of the corresponding linear fit.
    
    Note that Eq.~(\ref{eq_invpc}) can be rewritten as
    \begin{equation*}
    p_c=\frac{c}{z + b} = \frac{c}{z} \cdot \frac{1}{1 + \frac{b}{z}} = \frac{c}{z} - \frac{cb}{z^2} + \frac{cb^2}{z^3} - \frac{cb^3}{z^4} + \cdots,
    \end{equation*}
    \noindent therefore,
    \begin{equation*}
        zp_c = A + B\cdot\frac{1}{z} + \epsilon(z),
    \end{equation*}
    \noindent where $A = c$, $B = -cb$, and $\epsilon(z) = \mathcal{O}(1/z^2)$. For large $z$, the correction term $\epsilon(z)$ decays rapidly, and we obtain the approximate relation 
    \begin{equation}
    zp_c \approx A + B\cdot\frac{1}{z}.
    \label{zpc_inv_z}
    \end{equation}
    
    \noindent Therefore, plotting $z p_c$ versus $1/z$ provides a convenient way to assess convergence to the asymptotic regime. When focusing on the region where $z$ is sufficiently large, the intercept ($1/z \to 0$) yields an estimate of the limiting value of $z p_c$.

    \begin{figure}[t!]
    \begin{center}
        \includegraphics[width=10.0cm, height=10.0cm]{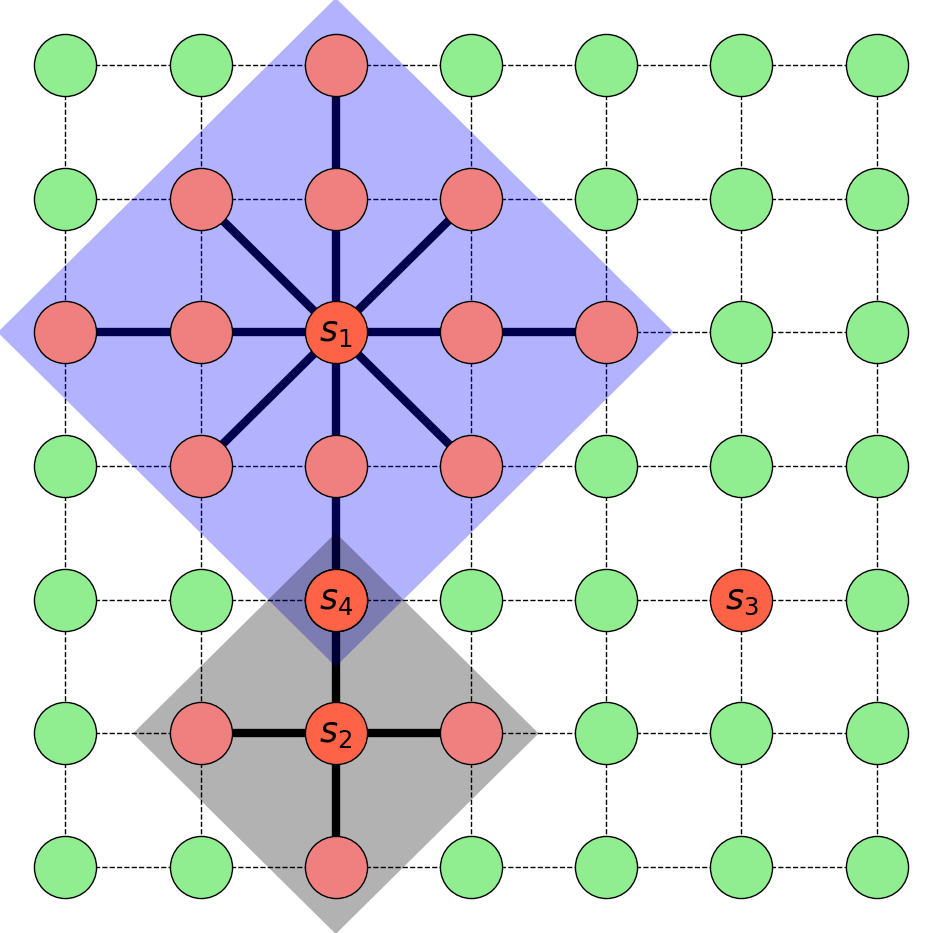} 
        \caption{Examples of compact diamond-shaped neighborhoods with radius $r$ for $r = 0$, $1$, and $2$. Sites $s_1$, $s_2$, $s_3$ and $s_4$ indicate the centers of the neighborhoods associated with each site. The blue and black diamonds illustrate the sites belonging to the neighborhood of $s_1$ (with $r = 2$) and $s_2$ (with $r = 1$), respectively. Note that $s_3$ and $s_4$ have $r = 0$, meaning that the diamonds centered at $s_3$ and $s_4$ have zero length. The figure also shows that although $s_4$ has $r = 0$, it lies within the neighborhoods of $s_1$ and $s_2$, implying that $s_1$, $s_4$, and $s_2$ are all connected.}
        \label{diamonds}
    \end{center}
    \end{figure}

	In this paper, we investigate a site percolation model, on the square lattice, with random extended neighborhoods in the form of compact diamond-shaped structures in a range of sizes. Assuming that the radius $r$ of a diamond is the maximum distance from its center to any site within it, this model is defined in the following steps:
	
	\begin{itemize}
		\item[(i)] Choose two non-negative integers, denoted as $i$ and $m$, where $i \leq m$.
		\item[(ii)] Consider a sequence of $i.i.d$ random variables $(r_s)_{s \in \mathbb{Z}^2}$ from a uniform distribution over the set $\{i, i+1, \ldots, m\}$.
		\item[(iii)] Define the neighborhood of site $s$ as a compact diamond-shaped region with radius $r_s$.
		\item[(iv)] Define percolation and the critical point in a manner consistent with the Bernoulli site percolation model.
	\end{itemize}
	
	Note that the probability of $r_s=j$, for any $j \in \{i, i+1, \ldots, m\}$, is $\rho_j=1/(m-i+1)$. We denote the critical point of the model by $p_c(i, m)$, and the average number of neighbors per site (\textit{average degree}) by $\overline{z}(i, m)$.

        It is important to note that two sites $s$ and $u$ are considered connected if either $s$ belongs to the neighborhood of $u$ or $u$ belongs to the neighborhood of $s$. For example, in Fig.\ \ref{diamonds}, we have $r_{s_4} = 0$, and site $s_4$ is connected to $s_1$ since it lies within the neighborhood of $s_1$. Furthermore, observe that even if only the sites $s_1$, $s_2$, and $s_4$ are open in the graph, they would still constitute a cluster of size 3.
        
\begin{figure}[t!]
    \begin{center}
        \includegraphics[width=9.5cm, height=9.5cm]{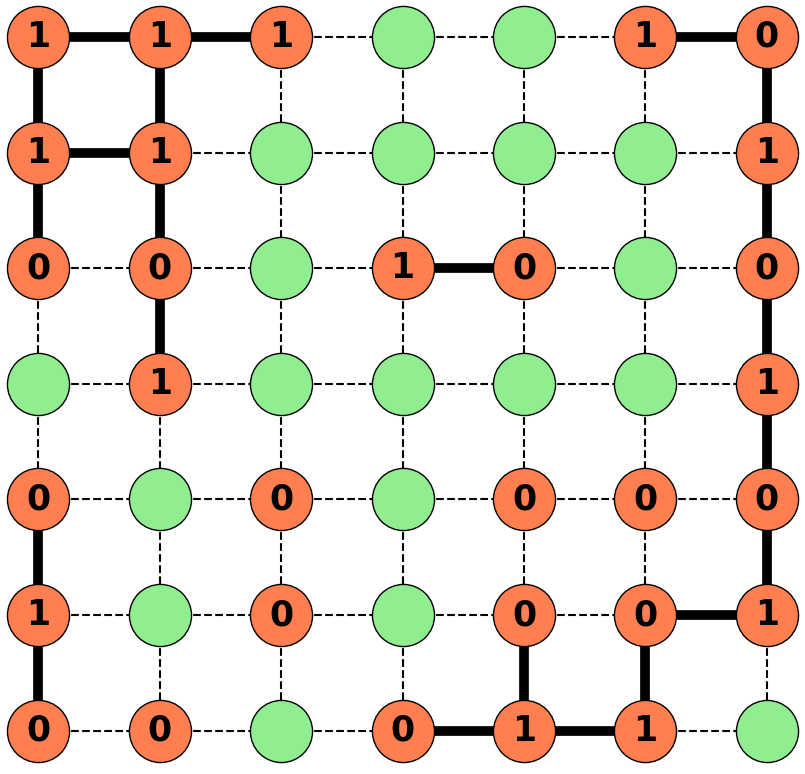} 
        \caption{Sample of the model for $i = 0$ and $m = 1$, where each site receives a value $r_s \in \{0, 1\}$ with equal probability. Open sites are shown in red and closed sites in green. The number inside each open site represents the associated value of $r_s$. When $r=1$, the site can connect with any of the four nearest neighbors, while when $r=0$ the site can act as a bridge. The black lines connect sites that belong to the same cluster. The figure illustrates how sites with $r_s = 0$ may be connected to other sites if they have adjacent sites with $r_s = 1$.}
        \label{01_case}
    \end{center}
    \end{figure}

        Fig.\ \ref{01_case} illustrates a sample of our model on a finite square lattice with free boundary conditions, considering $i = 0$ and $m = 1$ (each site has a neighborhood range of either $0$ or $1$), chosen with equal probability. Sites are shown in red if \textit{open}, and in green if \textit{closed}. The number inside each open site $s$ indicates the corresponding value of $r_s$. Bold solid lines between two sites indicate that they are connected and, therefore, belong to the same cluster. Note that some sites with $r_s = 0$ are connected to other sites and, therefore, may contribute to the occurrence of percolation.

    One possible application for this model lies in the study of disease spread using percolation models \cite{ziff}. In such a framework, an infection spreads between individuals and their contacts. An uncontrolled outbreak, where a large fraction of the population becomes infected, is then directly analogous to a percolation transition. For the model proposed here, the underlying assumption is that a larger neighborhood radius for an individual (site) represents greater mobility, and thus a higher transmission potential. For example, if $i=0$, the radius follows a uniform distribution over the set $\{0, 1, \ldots, m\}$, and smaller values of $m$ would correspond to stricter social distancing; conversely, larger values would imply less stringent policies.
             
	We studied this model by analyzing two different scenarios. In the first scenario, we considered the case where each site in the lattice has the same neighborhood radius, that is, $i = m$, and simulated the model for values of $m$ ranging from $1$ to $10$. Note that $m = 1$ corresponds to the classical Bernoulli site percolation model. We examined the behavior of the critical threshold $p_c(i,i)$ as a function of $\overline{z}(i,i)$ and, as expected, we observed the asymptotic behavior predicted by Eq.~\eqref{cont}. For simplicity of notation, we shall refer to $p_c(i,i)$ and $\overline{z}(i,i)$ simply as $p_c(i)$ and $\overline{z}(i)$ when discussing this case.

    In the second scenario, we investigated the behavior of $p_c(i,m)$ for fixed values of $i~\in~\{0, \dots, 7\}$ while varying $m$ from $i$ to $10$. In this case, each site is assigned a random neighborhood radius, uniformly drawn from the set $\{i, i+1, \dots, m\}$. For each $i$, we found that the average degree $\overline{z}(i,m)$ displays a linear relationship with $1/p_c(i,m)$, indicating a behavior analogous to that observed in the case $i=m$.

    For each fixed $i$, let us denote by $2^d\eta_c(i)$ (with $d=2$) the asymptotic value of ${\overline{z}(i,m) \cdot p_c(i,m)}$ as $m \to \infty$. The quantity $\eta_c(i)$ may be related to a continuum version of our model, in which points are added randomly to the continuum and two points are connected if either falls within the continuum neighborhood of the other. Such a model would differ from typical continuum percolation models, and we did not study it in this work. Our main observation is that $\overline{z}(i,m) \cdot p_c(i,m)$ appears to converge to a constant as $m \to \infty$ for fixed $i$.

    Given that in our model the neighborhood radius $r$ is uniformly distributed over $\{i, i+1, \ldots, m\}$, the relative influence of the lower bound $i$ on the value of $2^d\eta_c(i)$ becomes less significant as $i$ increases, since the range $\{i, i+1, \ldots, m\}$ shifts to larger values and the distribution becomes increasingly dominated by larger neighborhoods. The numerical results for $\overline{z}(i,m) \cdot p_c(i,m)$ shown in Table~\ref{table_pc_i_m} support this hypothesis. 
    
    Altogether, we simulated the model for $62$ distinct pairs $(i, m)$ and estimated their corresponding critical thresholds.
       		
    Gouker and Family \cite{gouker} studied this model when $i=m$ for $m=2,$ 4, 6, 8, and 10, and estimated the value of $p_c$ as given in Table \ref{table_pc_i_i}. Xun, Hao, and Ziff \cite{xun} investigated the site percolation model on square and simple cubic lattices, exploring various cases of extended neighborhoods, and among these cases, they considered the case where $i=m=2$ on the square lattice.

    Diamond-shaped objects are interesting as an example of a neighborhood that is of square shape and is an alternative to an aligned square on a square lattice.  How finite-size corrections compare for these two square boundaries is an interesting question to study --- is one orientation better than another?  Are there some special symmetries that favor one of the orientations? Since the critical threshold for the continuum percolation model, $\eta_c$, is expected to be the same for both diamond-shaped regions and aligned square regions on the square lattice, one might anticipate that the critical behavior would be comparable between the two orientations, particularly when analyzing the convergence of $z p_c$ to $\eta_c$.
    
    Using the results obtained by Jasna and Sasidevan~\cite{jasna2024} via the \textit{discrete excluded volume theory} for these two shapes, and applying relation~\eqref{cont}, we find that for aligned squares of side composed of $k$ sites,
    \begin{equation}
        z p_c \approx 4 \eta_c \cdot \frac{4k^2}{(2k+1)^2 - 5},
        \label{jasna_square}
    \end{equation}
    and for diamond-shaped squares with a diagonal composed of $k$ sites (corresponding to a radius equal to $(k - 1)/2$),
    \begin{equation}
        z p_c \approx 4 \eta_c \cdot \frac{k^2 + 1}{k(k + 1)}.
        \label{jasna_diamond}
    \end{equation}

    \noindent Therefore, the convergence rate of $z p_c$ to $\eta_c$ is essentially the same for both cases. For instance, when $k = 10$, the absolute difference between the two expressions is on the order of $7 \times 10^{-4}$.

\begin{figure}[h!]
		\begin{center}
			\includegraphics[width=9.5cm, height=9.5cm]{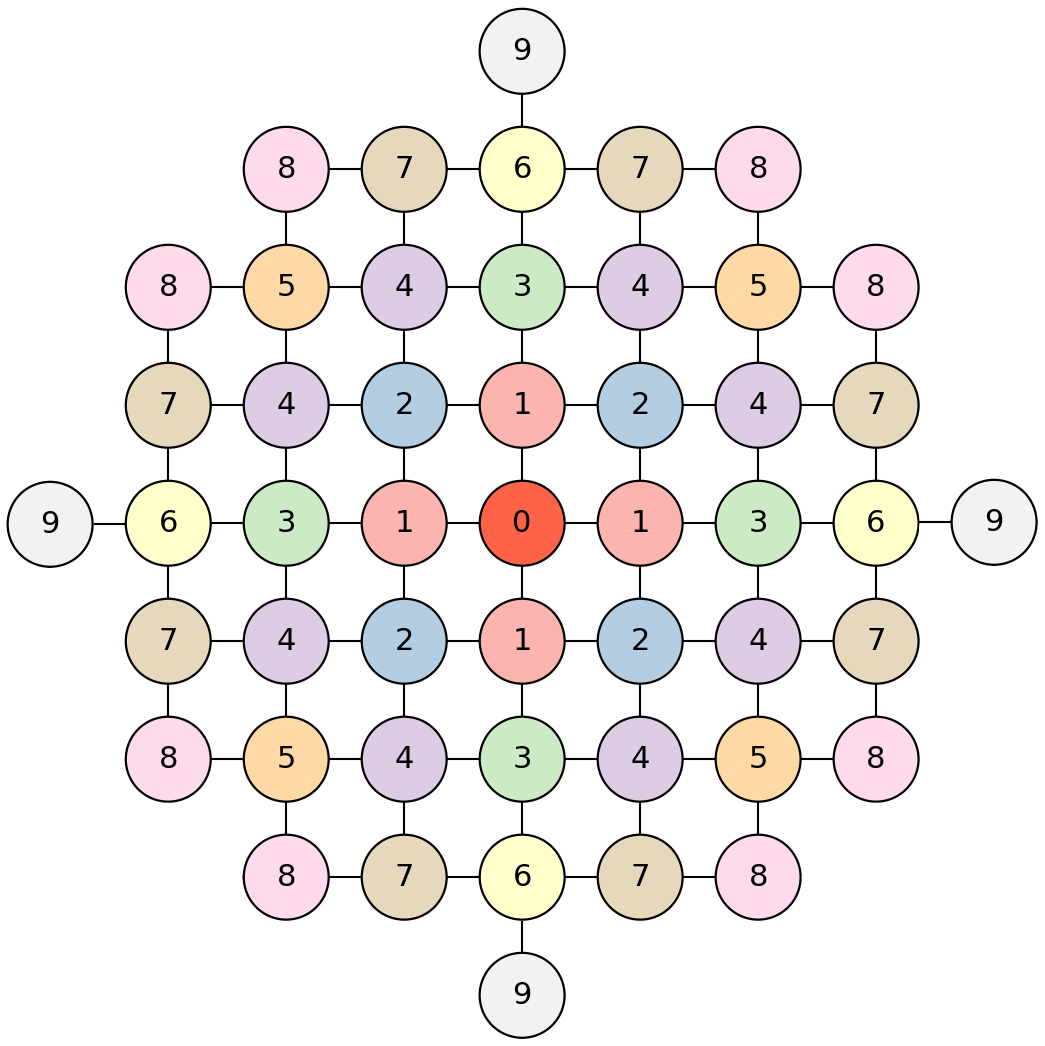} 
            \caption{Neighborhoods of a central site (labeled 0) on a square lattice, showing up to the ninth nearest neighbors. If a site is labeled with $n$, then it is the $n$th nearest neighbor of site 0. Note that a diamond of radius $r$ centered at each site corresponds to: the \textsc{sq}-1 neighborhood for $r=1$; the \textsc{sq}-1,2,3 neighborhood for $r=2$; and the \textsc{sq}-1,2,3,4,6 neighborhood for $r=3$.}
			\label{fig_sq}
		\end{center}
    \end{figure}

    We note that there has been a great deal of work on similar systems under the names ``equivalent-neighborhood models" \cite{dalton_1964, domb_1966, domb_1972, ouyang_2018}, ``extended-range percolation" \cite{xun, xun2, zhao_2022, cirigliano_2023, xun_2023, cirigliano_2024}, ``medium-range percolation" \cite{deng_2019}, ``range-$R$ percolation" \cite{frei_2016, hong_2023, hong_2025}, ``long-range percolation" \cite{gouker, iribarne_1999} and ``complex neighborhoods" \cite{majewski_2006, malarz_2022}.  The models here fall in the category of complex neighborhoods with connections to certain nearest neighbors.  For example, the $r=2$ model ($i=m=2$) corresponds to a square lattice with first nearest-neighbors, second nearest neighbors, and third nearest neighbors, described as square-NN+2NN+3NN or \textsc{sq}-1,2,3. Likewise $r = 3$ corresponds to \textsc{sq}-1,2,3,4,6, and $r = 4$ corresponds to \textsc{sq}-1,2,3,4,5,6,7,9. For larger $r$, the diamond neighborhoods do not correspond to fully filled NN shells. In Fig.~\ref{fig_sq}, we illustrate the $n$th nearest neighbors of a site for $1 \leq n \leq 9$. Several authors have looked at various similar neighborhoods, but of these only the \textsc{sq}-1,2,3 and \textsc{sq}-1,2,3,4,6 have been studied previously in Refs.~\cite{domb_1966, gouker,iribarne_1999,malarz_2005,majewski_2006,xun} ($p_c^{\text{site}} = 0.2891226(14)$ \cite{xun}) and in Ref.~\cite{malarz_2023} ($p_c^{\text{site}} = 0.16134$), respectively.

    \section{Numerical Procedure} \label{num_procedure}
	
	We consider periodic square lattices of lengths $L=2^j$, where $j$ ranged from $7$ to $11$, to be the set of vertices of the graph. For each pair $(i, m)$ we use the Newman-Ziff algorithm~\cite{ziff01} to estimate the critical threshold $p_c(i,m)$. The percolation criterion assumed is that the infinite cluster emerges when there exists a cluster that wraps around either the horizontal or vertical directions. The number of simulations varied from $2.5 \times 10^5$ ($L=2048$) to $10^7$ ($L=128$). This computational task consumed approximately two months of CPU time across 20 cores operating at a clock speed of 2.7 GHz.
	
	Once $L$, $i$ and $m$ are fixed, the percolation probability is given by
	
    \begin{equation}
    \psi_{_{\mbox{\scriptsize{$L$}}}}(p; i,m)=\sum_{j=0}^{N} 
    \binom{N}{j}
    p^{j} (1-p)^{N-j} \cdot  \overline{Q}_{j},
    \label{equal}
    \end{equation}
    \noindent where $p$ is the occupation probability, $ N $ is the total number of sites in the network, and $ \overline{Q}_{j} $ is the expected number of times percolation occurs when there are $j$ open sites. Each sample realization consists of a random sequence of all sites and, starting with all closed, we open them sequentially one at a time and then the percolation criterion is tested. Following the determination of $\psi_{_{\mbox{\scriptsize{$L$}}}}(p;i,m)$ for all values of $L$, the critical point $p_c(i,m)$ is estimated using the finite-size scaling 
    \begin{equation}
    [\overline{p}_{{_{\mbox{\tiny{$L$}}}}}(i,m)-p_{c}(i,m)] \propto L^{-{1}/{\nu}},
    \label{FSS}
    \end{equation}
    \noindent where 
    \begin{equation}
    \overline{p}_{{_{\mbox{\tiny{$L$}}}}}(i,m)=\int_{0}^{1} p \frac{d\psi_{_{\mbox{\tiny{$L$}}}}(p;i,m)}{dp} \ dp  \ = 1 - \int_{0}^{1} \psi_{_{\mbox{\tiny{$L$}}}}(p;i,m) \ dp
    \label{p_av_int}
    \end{equation}

    \noindent is defined as the \textit{average concentration} 
    at which percolation occurs for the first time and $\nu$ is the \textit{correlation length exponent} \cite{stauffer94}. In Ref.~\cite{gouker}, the authors found numerical evidence that when considering finite diamond-shaped structures, the model exhibits the same correlation-length exponent as the Bernoulli site percolation model ($\nu=\frac{4}{3}$), and it is generally expected that all short-ranged percolation models will obey this universal scaling behavior.

    As shown in Ref.\ \cite{ziff_newmann_2002}, one can substitute $\psi_L$ of (\ref{equal}) into (\ref{p_av_int}) and find that $\overline{p}_{{_{\mbox{\tiny{$L$}}}}}(i,m)$ can be found directly using the simple expression 
    \begin{equation}
        \overline{p}_{{_{\mbox{\tiny{$L$}}}}}(i,m)=\dfrac{1}{N+1}\sum_{n=0}^{N} n P_{\mbox{\tiny{$L$}},n},
        \label{p_av_direct}
    \end{equation}
    
    \noindent where $P_{\mbox{\tiny{$L$}},n}$ is the probability that the system first percolates when the number of occupied sites reaches $n$ \cite{ziff_newmann_2002}. In this way, it is not necessary to use the convolution in (\ref{equal}) and the integral expression in (\ref{p_av_int}). In our work, we estimate $p_c(i,m)$ using (\ref{p_av_direct}). We have also checked some values using (\ref{p_av_int}), and confirm that when the numerical integration interval $\Delta p$ decreases, the results of the two methods approach each other.

    In the compact diamond-shaped structures of radius $r$, each site has a number of neighbors given by $z_r=2r(r+1)$. It follows that the average degree $\overline{z}(i,m)$ is expressed as
    
    \begin{equation}
        \overline{z}(i,m)=\dfrac{1}{m-i+1} \sum_{j=i}^{m} z_j = \dfrac{2}{3}(i^2+mi+i+m^2+2m)
    \end{equation}
    
    \noindent since the neighborhood size distribution of each site follows a uniform distribution in $\{i, i+1, ..., m\}$.

	Following the estimation of $p_c(i,m)$ for all cases, we proceeded to analyze the asymptotic behavior of $\overline{z}(i,m) \cdot p_c(i,m)$. This analysis was conducted across the two scenarios studied: (i) where $i = m$ for $i \in \{1, \ldots, 10\}$; and (ii) where $i$ was held constant ($0 \leq i \leq 7$) while $m$ varied from $i$ to $10$. For each scenario, the two theoretical approaches outlined in Section 1, corresponding to relations \eqref{eq_invpc} and \eqref{zpc_inv_z}, were employed.

    Linear regressions were performed using $\overline{z}(i,m)$ \textit{versus} $1/p_c(i,m)$ (corresponding to Eq.~\eqref{eq_invpc}), as well as $\overline{z}(i,m)\cdot p_c(i,m)$ \textit{versus} $1/\overline{z}(i,m)$ (corresponding to Eq.~\eqref{zpc_inv_z}). In each case, the slope and intercept were estimated. For the first relation, the slope provides an estimate of $2^d\eta_c(i)$ in the case of random neighborhood sizes and of $2^d\eta_c$ in the case of identical neighborhood sizes, while for the second relation, this estimate is given by the intercept.

    Critical thresholds and slope uncertainties were obtained from the standard deviation of the regression residuals.

    For the case in which sites have only neighborhood radius $0$ and $1$, we also analyzed how the critical threshold behaves by considering a fraction $\rho_0$ of sites with neighborhood radius 0 and $\rho_1$ with size 1 ($\rho_0 + \rho_1 = 1$). This setup allows us to assess the impact of sites with neighborhood size $0$ on the occurrence of percolation. In this scenario, we simulated the model just for $L = 1024$ using $10^5$ samples.

   \section{Results}

\begin{figure}[h!]
		\begin{center}
			\includegraphics[width=14cm, height=9.0cm]{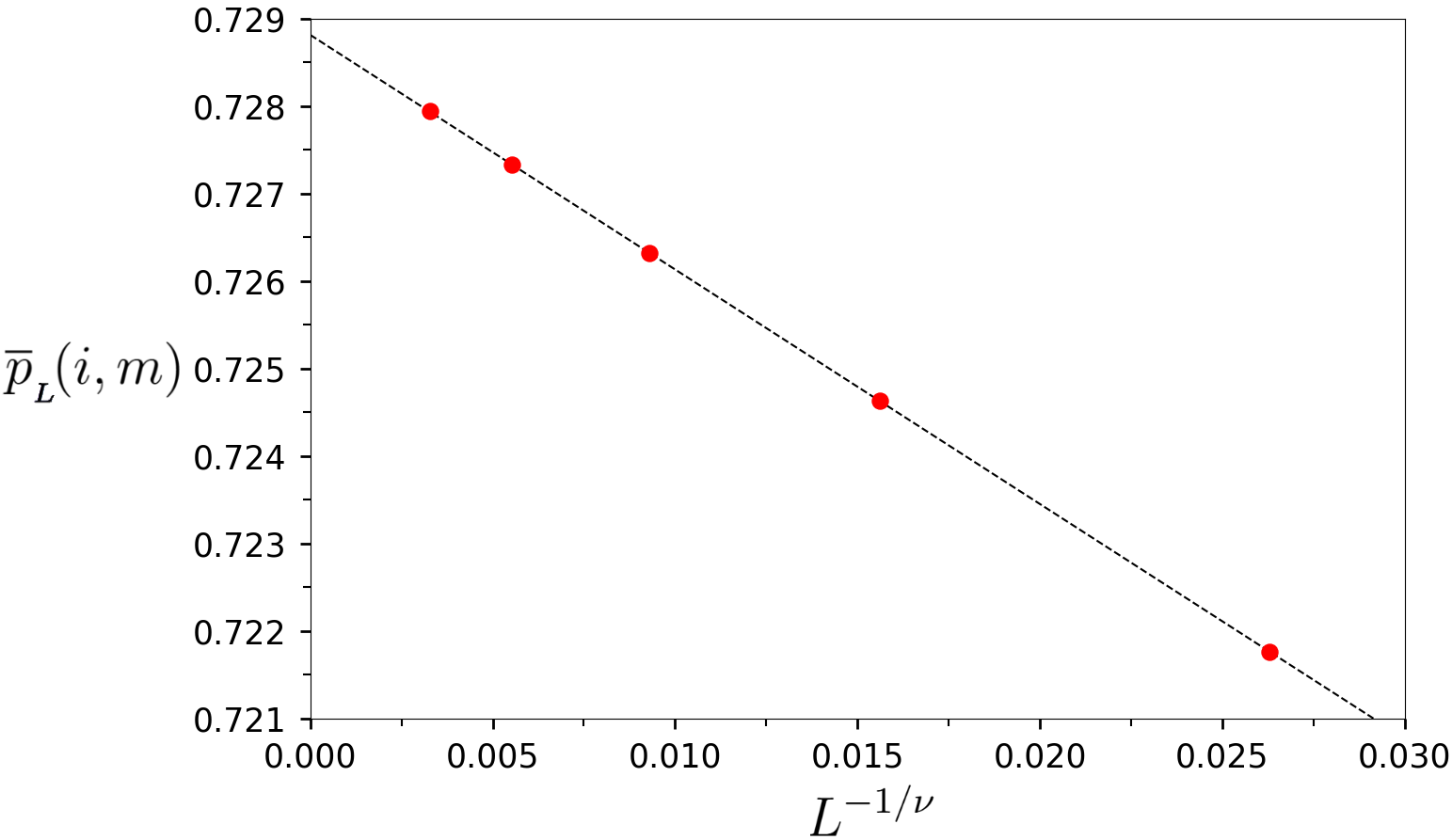} 
            \caption{Estimate of the critical point $p_c(i,m)$ for the case $(i,m) = (0,1)$ using the finite-size scaling relation~\eqref{FSS}. The intercept of the fitted line with the vertical axis gives the estimate $p_c(0,1) = 0.72883(1)$. Similar plots were obtained for other values of $(i,m)$, and are therefore omitted.}
			\label{fig_FSS}
		\end{center}
    \end{figure}

    For each pair $(i, m)$ analyzed, the critical threshold $p_c(i, m)$ was estimated using the finite-size scaling relation given in Eq.~\eqref{FSS}. Fig.\ \ref{fig_FSS} shows the plot of $\overline{p}_{\mbox{\tiny{$L$}}}(i, m)$ \textit{versus} $L^{-1/\nu}$ for the case $i = 0$ and $m = 1$, which corresponds to a system with equal number of sites with zero nearest-neighbors (which can act as bridges) and one nearest neighbor. The fit includes data from lattice sizes $L = 128$, $256$, $512$, $1024$, and $2048$, assuming $\nu = 4/3$. The intercept of the linear fit yields the estimated critical threshold for this case: $p_c(0, 1) = 0.72883(1)$.

    Table~\ref{table_pc_i_m} presents all estimated values of $p_c(i,m)$ for $0 \leq i \leq 7$ and $i \leq m \leq 10$. Table~\ref{table_pc_i_i} shows the corresponding estimates for $p_c(i)$, with $1 \leq i \leq 10$. In addition to the critical thresholds, we also report the values of the average degree $\overline{z}(i,m)$ and the product $\overline{z}(i,m) \cdot p_c(i,m)$. In all cases, the results are consistent with the convergence of this product to a constant value.

    \begin{table}[h!]
    \renewcommand{\arraystretch}{0.8}
    \setlength{\extrarowheight}{0.2cm}
    \setlength{\tabcolsep}{2.1pt}
    \scriptsize
    \begin{tabular}{|c|c|c|c|c|c|c|c|c|c|c|}
        \cline{2-11}
         \multicolumn{1}{c|}{} & \multicolumn{1}{c|}{\textbf{$m=1$}} & \textbf{$m=2$} & \textbf{$m=3$} & \textbf{$m=4$} & \textbf{$m=5$} & \textbf{$m=6$} & \textbf{$m=7$} & \textbf{$m=8$} & \textbf{$m=9$} & \textbf{$m=10$} \\ 
        \cline{2-11}
        \hline
        \multicolumn{1}{|c|}{$p_c(0,m)$} & 0.72883(1) & 0.39134(1) & 0.23581(1) & 0.15559(1) & 0.10983(1) & 0.08148(1) & 0.06277(1) & 0.04981(1) & 0.04047(1) & 0.03352(1) \\
        \textbf{$\overline{z}(0,m)$} & 2.0 & 5.$\bar{3}$ & 10.0 & 16.0 & 23.$\bar{3}$ & 32.0 & 42.0 & 53.$\bar{3}$ & 66.0 & 80.0 \\
        \textbf{$\overline{z}\cdot p_c$} & 1.458 & 2.087 & 2.358 & 2.489 & 2.563 & 2.607 & 2.636 & 2.657 & 2.671 & 2.682 \\
        \hline
        \hline
        \textbf{$p_c(1,m)$} & 0.59275(1) & 0.33235(1) & 0.20726(1) & 0.14007(1) & 0.10057(1) & 0.07554(1) & 0.05875(1) & 0.04696(1) & 0.03838(1) & 0.03194(1) \\
        \textbf{$\overline{z}(1,m)$} & 4.0 & 8.0 & 13.$\bar{3}$ & 20.0 & 28.0 & 37.$\bar{3}$ & 48.0 & 60.0 & 73.$\bar{3}$ & 88\\
        \textbf{$\overline{z}\cdot p_c$} & 2.371 & 2.659 & 2.763 & 2.801 & 2.816 & 2.814 & 2.820 & 2.818 & 2.814 & 2.811 \\
        \hline
        \hline
        \textbf{$p_c(2,m)$} & -- & 0.28912(1) & 0.18094(1) & 0.12516(1) & 0.09152(1) & 0.06969(1) & 0.05477(1) & 0.04414(1) & 0.03630(1) & 0.03038(1) \\
        \textbf{$\overline{z}(2,m)$} & -- & 12.0 & 18.0 & 25.$\bar{3}$ & 34.0 & 44.0 & 55.$\bar{3}$ & 68.0 & 82.0 & 97.$\bar{3}$ \\
        \textbf{$\overline{z}\cdot p_c$} & -- & 3.469 & 3.257 & 3.167 & 3.112 & 3.066 & 3.031 & 3.002 & 2.977 & 2.955 \\
        \hline
        \hline
        \textbf{$p_c(3,m)$} & -- & -- & 0.16132(1) & 0.11169(1) & 0.08297(1) & 0.06405(1) & 0.05088(1) & 0.04136(1) & 0.03426(1) & 0.02882(1) \\
        \textbf{$\overline{z}(3,m)$} & -- & -- & 24.0 & 32.0 & 41.$\bar{3}$ & 52.0 & 64.0 & 77.$\bar{3}$ & 92.0 & 108.0 \\
        \textbf{$\overline{z}\cdot p_c$} & -- & -- & 3.872 & 3.574 & 3.427 & 3.331 & 3.256 & 3.195 & 3.152 & 3.113 \\
        \hline
        \hline
        \textbf{$p_c(4,m)$} & -- & -- & -- & 0.10178(1) & 0.07527(1) & 0.05874(1) & 0.04716(1) & 0.03866(1) & 0.03226(1) & 0.02730(1) \\
        \textbf{$\overline{z}(4,m)$} & -- & -- & -- & 40.0 & 50.0 & 61.$\bar{3}$ & 74.0 & 88.0 & 103.$\bar{3}$ & 120.0  \\
        \textbf{$\overline{z}\cdot p_c$} & -- & -- & -- & 4.071 & 3.764 & 3.599 & 3.490 & 3.402 & 3.332 & 3.276 \\
        \hline
        \hline
        \textbf{$p_c(5,m)$} & -- & -- & -- & -- & 0.06952(1) & 0.05398(1) & 0.04367(1) & 0.03609(1) & 0.03032(1) & 0.02582(1) \\
        \textbf{$\overline{z}(5,m)$} & -- & -- & -- & -- & 60.0 & 72.0 & 85.$\bar{3}$ & 100.0 & 116.0 & 133.$\bar{3}$\\
        \textbf{$\overline{z}\cdot p_c$} & -- & -- & -- & -- & 4.171 & 3.887 & 3.726 & 3.609 & 3.517 & 3.439 \\
        \hline
        \hline
        \textbf{$p_c(6,m)$} & -- & -- & -- & -- & -- & 0.05041(1) & 0.04053(1) & 0.03369(1) & 0.02848(1) & 0.02439(1) \\
        \textbf{$\overline{z}(6,m)$} & -- & -- & -- & -- & -- & 84.0 & 98.0 & 113.$\bar{3}$ & 130.0 & 148.0\\
        \textbf{$\overline{z}\cdot p_c$} & -- & -- & -- & -- & -- & 4.234 & 3.972 & 3.816 & 3.702 & 3.610 \\
        \hline
        \hline
        \textbf{$p_c(7,m)$} & -- & -- & -- & -- & -- & -- & 0.03814(2) & 0.03151(1) & 0.02675(1) & 0.02303(1) \\
        \textbf{$\overline{z}(7,m)$} & -- & -- & -- & -- & -- & -- & 112.0 & 128.0 & 145.$\bar{3}$ & 164.0\\
        \textbf{$\overline{z}\cdot p_c$} & -- & -- & -- & -- & -- & -- & 4.272 & 4.033 & 3.886 & 3.777 \\
        \hline
     \end{tabular}
    \caption{Estimated values of the critical threshold $p_c(i,m)$ for each pair $(i,m)$, with $i \in \{0,...,7\}$ and $i\leq m \leq 10$, along with the corresponding average degree $\overline{z}(i,m)$ and product $\overline{z} \cdot p_c$. The values of $p_c(i,m)$ are obtained via finite-size scaling (\ref{FSS}). Note that the overline indicates a repeating decimal (e.g., $5.\bar{3} = 5.333\ldots$).}
    \label{table_pc_i_m}
\end{table}

\begin{table}[h!]
    \renewcommand{\arraystretch}{0.95}
    \setlength{\extrarowheight}{0.2cm}
    \setlength{\tabcolsep}{4.5pt}
    \small
    \centering
    \begin{tabular}{|c|c|c|c||c|c|}
        \cline{3-6}
        \multicolumn{1}{c}{} & \multicolumn{1}{c}{} & \multicolumn{2}{|c||}{our study} & \multicolumn{2}{|c|}{previous studies}   \\
        \hline
        $i$ & $\overline{z}(i)$ & $p_c(i)$ & $\overline{z} \cdot p_c$ & $p_c(i)$ - Ref.\cite{gouker} & $p_c(i)$ - Ref.\cite{jasna2024} \\
        \hline
        2  & 12  & 0.28912(1) & 3.469 & 0.290(5) & 0.31744 \\
        3  & 24  & 0.16132(1) & 3.872 & -        & 0.16352 \\
        4  & 40  & 0.10178(1) & 4.071 & 0.105(5) & 0.10012 \\
        5  & 60  & 0.06952(1) & 4.171 & -        & 0.06771 \\
        6  & 84  & 0.05041(1) & 4.234 & 0.049(5) & 0.04888 \\
        7  & 112 & 0.03814(1) & 4.272 & -        & 0.03696 \\
        8  & 144 & 0.02986(1) & 4.300 & 0.028(5) & 0.02893 \\
        9  & 180 & 0.02398(1) & 4.316 & -        & 0.02326 \\
        10 & 220 & 0.01969(1) & 4.332 & 0.019(5) & 0.01911 \\
        \hline
    \end{tabular}
    \caption{Comparison of the critical percolation thresholds $p_c(i)$ obtained in our study with previous estimates from the literature, for $2 \leq i \leq 10$. We also report the corresponding average degrees $\overline{z}(i)$ and the products $\overline{z}(i) \cdot p_c(i)$. The values from Ref.~\cite{jasna2024} are analytical estimates based on excluded volume theory and become more accurate as $i$ increases.}
    \label{table_pc_i_i}
    \end{table}

    For clarity of presentation, we organize the remainder of the results into two parts. In the first, we examine the case $i = m$, where all sites have neighborhoods of the same size. In the second, we consider the case where each site has a randomly assigned neighborhood radius, drawn uniformly from the set $\{i, i+1, \dots, m\}$; this scenario is analyzed for all pairs $(i, m)$ with $i \in \{0, \dots, 7\}$ and $i \leq m \leq 10$.

    \subsection{Identical neighborhood sizes}

    We include in Table~\ref{table_pc_i_i} values of $p_c(i)$ reported by Gouker and Family~\cite{gouker}, based on Monte Carlo simulations, and by Jasna and Sasidevan~\cite{jasna2024}, who applied the excluded volume theory adapted to a lattice setting~\cite{koza}. The theoretical estimates from~\cite{jasna2024} are based on an analytical approximation derived from the excluded volume theory and are expected to converge toward the true values of $p_c(i)$ as the diamond size increases. As shown in Table~\ref{table_pc_i_i}, their estimates become increasingly closer to our numerical results for larger values of $i$. When $i=2$, several additional estimates of $p_c(i)$ can be found in the literature: $0.292$~\cite{domb_1966}, $0.290(5)$~\cite{gouker}, $0.289$~\cite{iribarne_1999}, $0.288$~\cite{malarz_2005,majewski_2006}, and $0.2891226(14)$~\cite{xun}. Our result, $p_c(2) = 0.28912(1)$, is consistent with the more precise estimate $p_c = 0.2891226(14)$ reported in~\cite{xun}. For $i = 3$, Malarz~\cite{malarz_2023} reported an estimate of $p_c(3) = 0.16134$; although no uncertainty was provided, our result, $p_c(3) = 0.16132(1)$, is likely compatible with his, given the numerical proximity.

    The values of the product $\overline{z}(i) \cdot p_c(i)$ reported in Table~\ref{table_pc_i_i} exhibit a trend consistent with convergence toward the theoretical prediction $2^d \eta_c \approx 4.395$, with $d = 2$ and $\eta_c=1.09884280(9)$ \cite{mertens}. To analyze the asymptotic behavior of $\overline{z}(i) \cdot p_c(i)$, we considered two approaches based on the scaling relations discussed in Section~\ref{num_procedure}. 

\begin{figure}[h!]
	\begin{center}
		\includegraphics[width=11.0cm, height=9cm]{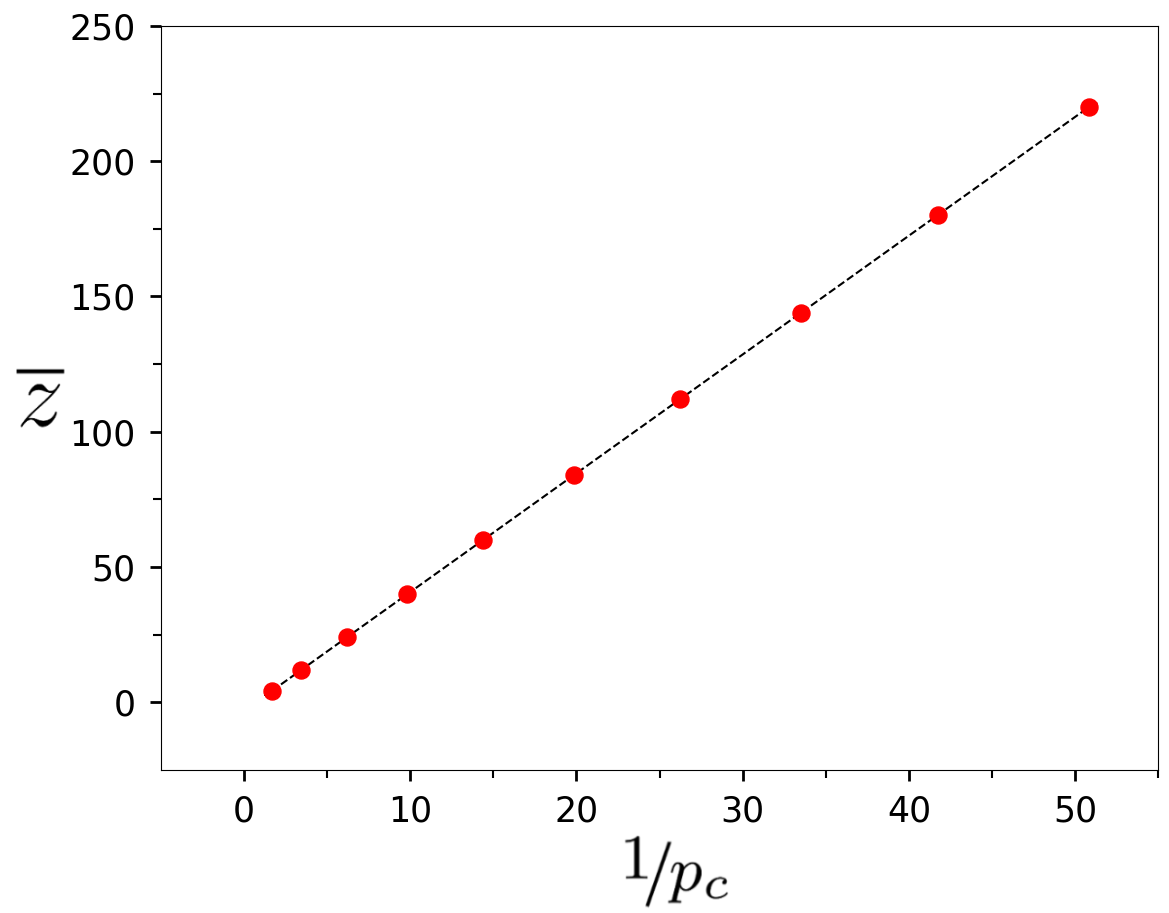} 
        \caption{Graph illustrating the relationship between $1/p_c(i)$ and $\overline{z}(i)$ for the cases where $i = m$ ($1 \leq i \leq 10$), that is, all sites have neighborhoods with fixed radius $r = i$. The figure shows that the dependence between the average number of neighbors $\overline{z}$ and the inverse of the critical point is approximately linear. The black dashed line is a fit through the data, and has a slope $4.396(2)$, consistent with the predicted value $4.395$ from Eq.\ (\ref{cont}).}
		\label{fig_equal}
	\end{center}
    \end{figure}

    First, we plotted $\overline{z}(i)$ \textit{versus} $1/p_c(i)$ (Fig.\ \ref{fig_equal}), corresponding to Eq.~\eqref{eq_invpc}. The slope obtained from the linear regression yields an estimate of $2^d \eta_c$ equal to $4.396(2)$, in excellent agreement with the theoretical prediction of $2^d \eta_c \approx 4.395$.

\begin{figure}[httb]
	\begin{center}
		\includegraphics[width=12cm, height=9cm]{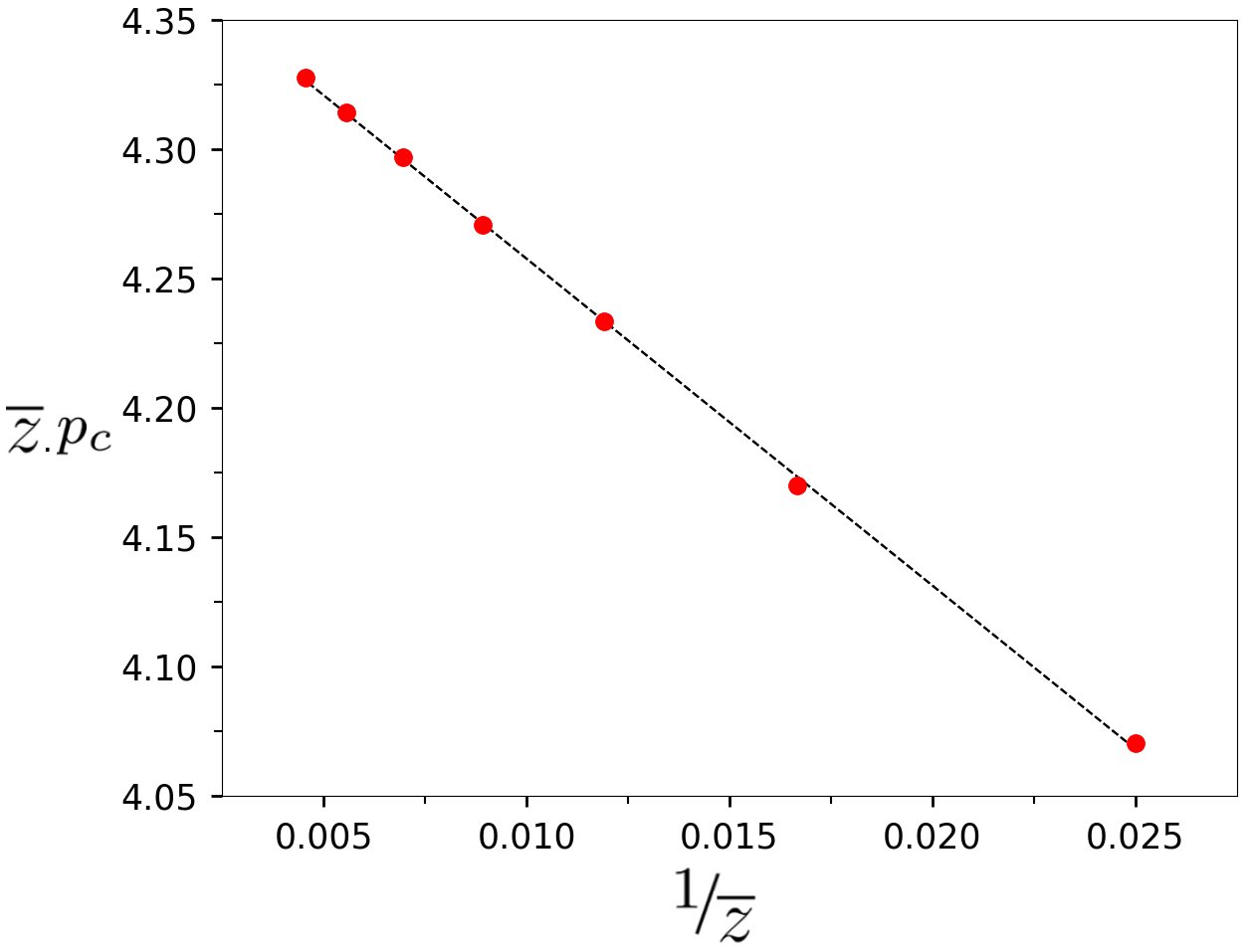} 
        \caption{Graph illustrating the relationship between $\overline{z}(i) \cdot p_c(i)$ and $1/z(i)$, with $i \in \{1, 2, \ldots, 10\}$. As in Fig.\ \ref{fig_zpc_i_neq_m}, the goal is to assess whether the product $\overline{z} \cdot p_c$ tends toward a constant as the average degree $\overline{z}$ increases. A linear fit is applied to the data for $\overline{z}(i) \geq 40$, yielding a slope of $-12.95(8)$ and an intercept of $4.387(1)$. This intercept estimates the limiting value of $\overline{z} \cdot p_c$ in the regime of large connectivity.}
		\label{fig_zpc_i_equal_m}
	\end{center}
    \end{figure}

    Second, we analyzed the relation $\overline{z}(i) \cdot p_c(i)$ \textit{versus} $1/\overline{z}(i)$ (Fig.\ \ref{fig_zpc_i_equal_m}), associated with Eq.~\eqref{zpc_inv_z}. Since this approximation is valid in the large-$z$ regime, we restricted the analysis to $\overline{z}(i) \geq 40$. The linear fit yields an intercept of $4.387(1)$, which, although slightly lower than the previous estimate, remains close to the expected value of $2^d \eta_c$.

    In both approaches, the data points exhibit a clear linear trend, further supporting the validity of the scaling relations~\eqref{eq_invpc} and \eqref{zpc_inv_z} in the case of identical neighborhood sizes.
    
    \subsection{Random neighborhood sizes}

    For the case where each site has a neighborhood radius uniformly drawn from $\{i, i+1, \ldots, m\}$, we analyzed the asymptotic behavior of the product $\overline{z}(i,m) p_c(i,m) \approx 2^{d} \eta_c(i)$, where $d = 2$ and $\eta_c(i)$ does not seem to have a direct correspondence with the critical threshold of a typical continuum percolation model, unlike the monodisperse case ($i = m$), where $\overline{z}(i)\,p_c(i)$ tends to $2^d \eta_c$.
    
\begin{figure}[h!]
		\begin{center}
			\includegraphics[width=14.5cm, height=10.0cm]{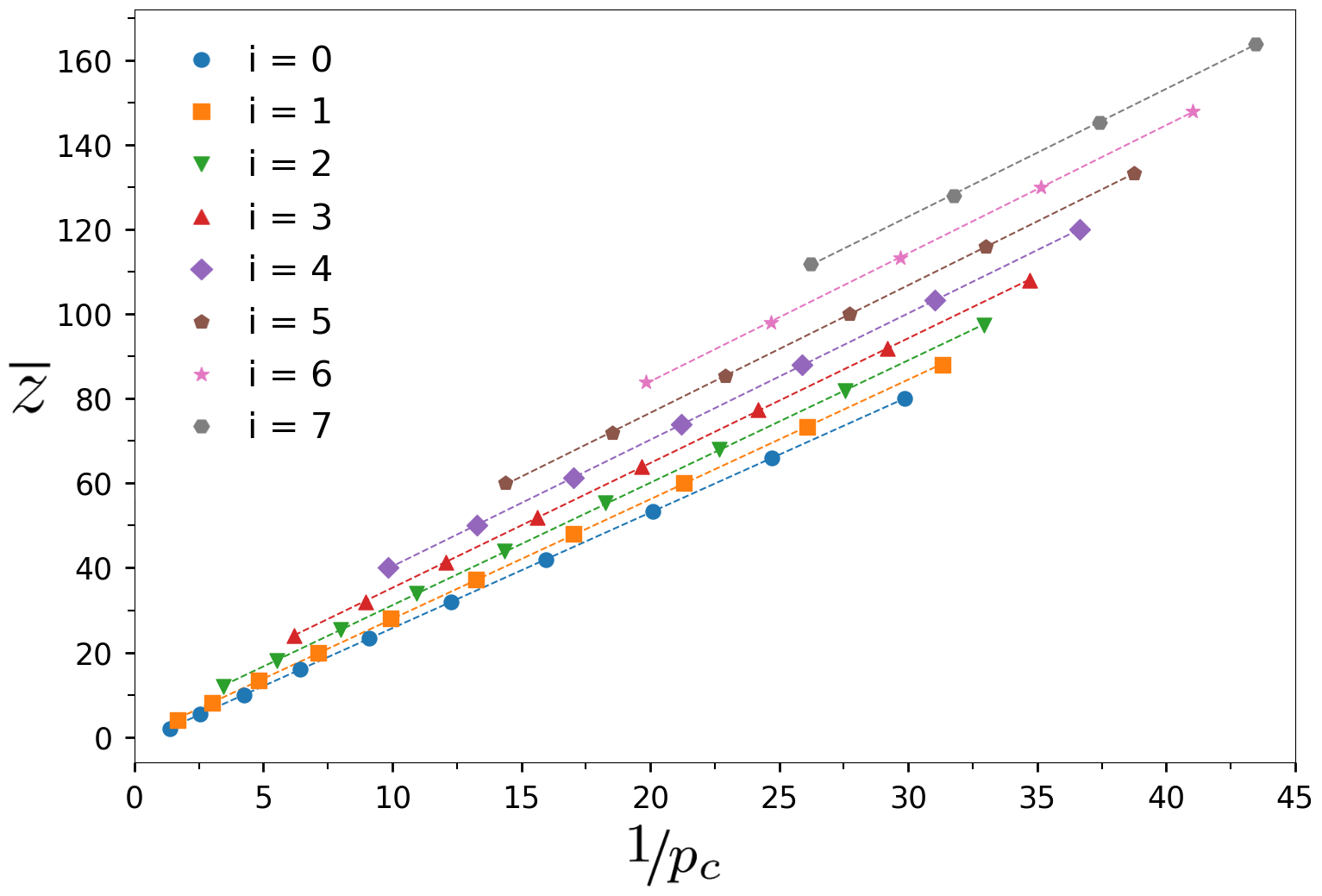} 
            \caption{Graphs illustrating the relationship between $\overline{z}(i,m)$ and $1/p_c(i,m)$ for different fixed values of $i \in \{0,1,\ldots,7\}$, with $m$ ranging from $i$ to $10$. Each set of points corresponds to a fixed value of $i$ and shows how the inverse of the critical point behaves as the average number of neighbors $\overline{z}(i,m)$ increases. The neighborhood radius distribution is uniform over $\{i, i+1, \ldots, m\}$. The dashed lines are a fit through the data; the slope and the intercept of each one are provided in Table~\ref{table_eta_c_i}.}
			\label{fig_all}
		\end{center}
	\end{figure}

\begin{figure}[h!]
		\begin{center}
			\includegraphics[width=14.5cm, height=10cm]{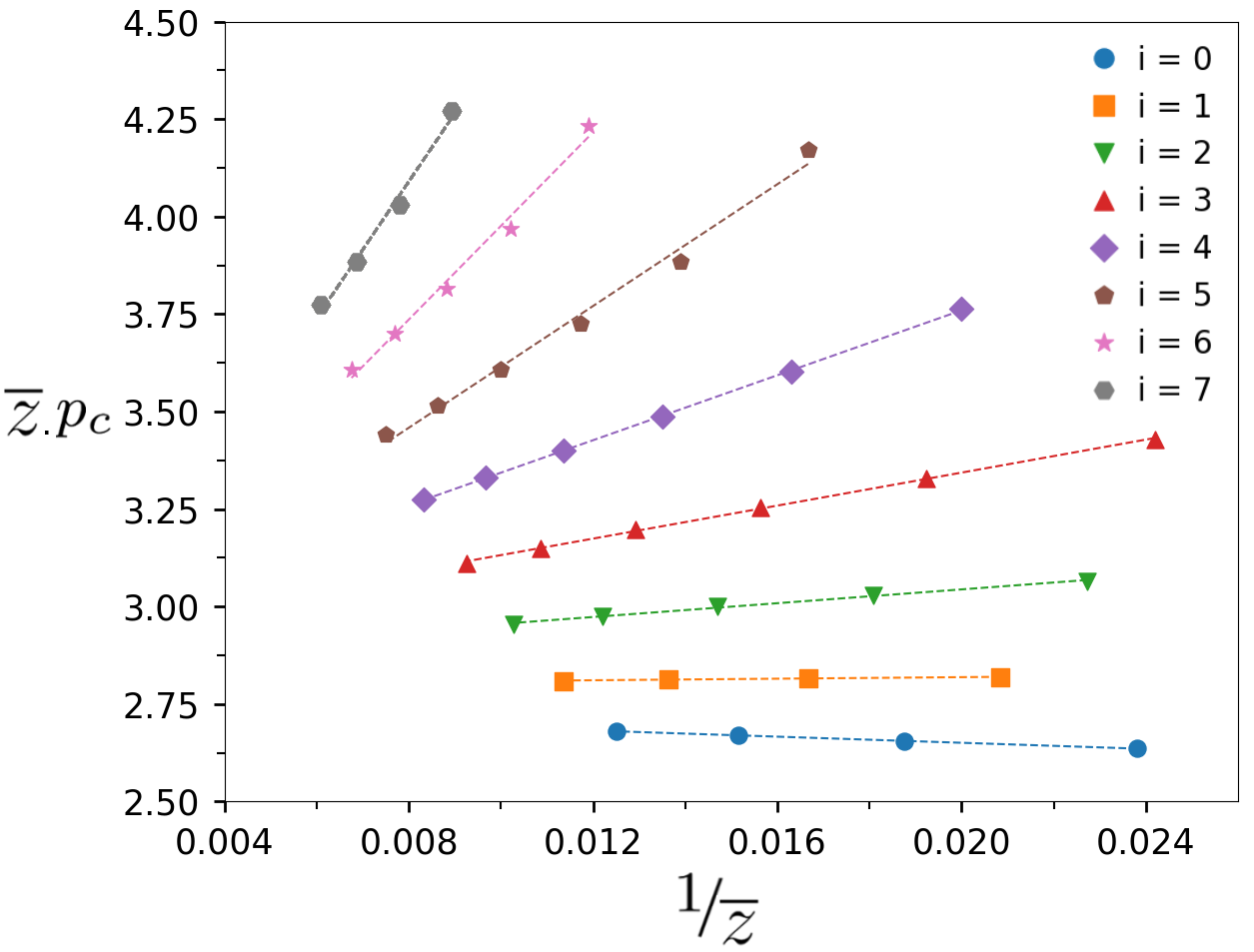} 
            \caption{Graphs illustrating the relationship between $\overline{z}(i,m) \cdot p_c(i,m)$ and $1/\overline{z}(i,m)$ for different fixed values of $i \in \{0,1,\ldots,7\}$, with $m$ ranging from $i$ to $10$. Each set of points corresponds to a fixed value of $i$. Linear behavior is expected for high $\overline{z}(i,m)$ values, so we only consider pairs $(i,m)$ where $\overline{z}(i,m) > 40$. This analysis aims to verify if the product $\overline{z} \cdot p_c$ approaches a constant in the high connectivity limit, as predicted by continuum percolation approximations. Table~\ref{table_eta_c_i} presents the values of the slopes and intercepts of the graphs. Note that the intercept refers to the value of $\overline{z} \cdot p_c$ when $\overline{z} \to \infty$.}
			\label{fig_zpc_i_neq_m}
		\end{center}
    \end{figure}

    This was done by fixing $i \in \{0, \ldots, 7\}$, computing the values of $\overline{z}(i,m) \cdot p_c(i,m)$ for $m$ ranging from $i$ to $10$, and extrapolating the results to $m \to \infty$ using the two scaling relations~\eqref{eq_invpc} and \eqref{zpc_inv_z}. The estimated values of $2^d \eta_c(i)$ obtained from the corresponding linear regressions are presented in Table~\ref{table_eta_c_i}. In addition, Figs.~\ref{fig_all} and~\ref{fig_zpc_i_neq_m} display the data and fitted lines.

    \begin{table}[h!]
		\renewcommand{\arraystretch}{1.0} 
		\setlength{\extrarowheight}{0.25cm} 
		\setlength{\tabcolsep}{3.5pt} 
		\centering
        \small
		\begin{tabular}{cc|c|c|c|c|c|c|c|c|}
			\cline{3-10}  
			& \multicolumn{1}{c|}{} & \multicolumn{1}{c|}{$i=0$} & \multicolumn{1}{c|}{$i=1$} & \multicolumn{1}{c|}{$i=2$} & \multicolumn{1}{c|}{$i=3$} & \multicolumn{1}{c|}{$i=4$} & \multicolumn{1}{c|}{$i=5$} & \multicolumn{1}{c|}{$i=6$} & \multicolumn{1}{c|}{$i=7$} \\ \cline{2-7} \hline
			\multicolumn{1}{|c|}{\multirow{2}{*}{$2^d\eta_c(i)$}} & \multicolumn{1}{c|}{Eq. \ref{eq_invpc}}    & 2.738(2)        & 2.831(8)        & 2.899(7)         & 2.954(7)        & 2.994(6) 	& 3.022(10)  	& 3.034(18)  & 3.023(31)   \\ 
           \multicolumn{1}{|c|}{}     & \multicolumn{1}{c|}{Eq. \ref{zpc_inv_z}} & 2.731(1)       & 2.801(2)      & 2.869(5)        & 2.923(5)       & 2.929(4)	& 2.837(45)   & 2.779(66)   & 2.695(92)     \\  \hline
		\end{tabular}
        \caption{Estimated values of $2^d \eta_c(i)$ for $i \in \{0, \dots, 7\}$. Each row corresponds to a specific fitting procedure: the first row uses the slope from the linear fit of $\overline{z}(i,m)$ versus $1/p_c(i,m)$ (Eq.~\eqref{eq_invpc}, Fig.\ \ref{fig_all}); and the second row shows the intercept from the fit of $\overline{z}(i,m)\cdot p_c(i,m)$ versus $1/\overline{z}(i,m)$ (Eq.~\eqref{zpc_inv_z}, Fig.\ \ref{fig_zpc_i_neq_m}). All fits were performed for fixed values of $i$ with $m$ ranging from $i$ to $10$.}  
        \label{table_eta_c_i}
    \end{table}

    In both cases, the scaling relations yielded data points that align well with a linear fit, suggesting that the assumptions behind Eqs.~\eqref{eq_invpc} and \eqref{zpc_inv_z} are valid even in the setting of random neighborhood sizes. 

    When comparing the results across the two approaches, we find that the estimates of $2^d \eta_c(i)$ obtained via Eq.~\eqref{zpc_inv_z} (intercept of $\overline{z} \cdot p_c$ versus $1/\overline{z}$) tend to be slightly lower than those from Eq.~\eqref{eq_invpc}, especially for larger values of $i$. This difference can be attributed to the fact that plots of $\overline z p_c$ vs.\ $1/\overline z$ (Fig.\ \ref{fig_zpc_i_neq_m}) are not very linear.  In other words, this form is not a good one to capture higher-order corrections well.

    Finally, we observe that as $i$ increases, the values of $2^d \eta_c(i)$ become increasingly stable. For instance, in the first two rows of Table~\ref{table_eta_c_i}, the estimates for $i=5$, $i=6$, and $i=7$ differ by less than $0.02$, which supports the hypothesis that the influence of the lower bound $i$ in the uniform distribution $\{i, i+1, \ldots, m\}$ becomes negligible as $i$ grows. 

    These results also highlight that the critical behavior of the system is influenced not only by the average neighborhood size but also by the specific distribution of neighborhood sizes. Systems with similar average degree $\overline{z}$ but different values of $i$ and $m$ can display significantly different percolation thresholds $p_c$. For example, the case $(i,m) = (2,3)$ yields $\overline{z} = 18$ and $p_c = 0.18094(1)$, while for $(i,m) = (0,4)$ we have $\overline{z} = 16$ and $p_c = 0.15559(1)$. Despite having a lower average degree, the latter case exhibits a significantly smaller critical threshold. This behavior can be attributed to the broader distribution of neighborhood radius in the $(0,4)$ model, which allows the system to establish long-range connections and thereby facilitates percolation.
    
    This comparison suggests that systems with a wider spread in neighborhood radius tend to percolate at lower occupation probabilities than those with narrower or fixed-size neighborhoods, even when the average degree is comparable.

    To assess orientation effects, we compare the results for aligned squares with those for diamond-shaped regions having the same average coordination number $\overline{z}$. First, note that for a diamond of radius $r$, there are $k = 2r + 1$ sites along the diagonal, and the average degree is $\overline{z} = (k^2 - 1)/2$. For aligned squares with side length composed of $l$ sites, the average degree is $\overline{z} = l^2 - 1$.

    There are infinitely many pairs $(k, l)$ that yield diamond and square configurations with the same average degree. This can be shown by observing that the condition
    \begin{equation}
    \frac{1}{2}(k^2 - 1) = l^2 - 1
    \label{eq_sq_diam}
    \end{equation}
    is equivalent to the negative Pell equation $k^2 - 2l^2 = -1$, which admits infinitely many integer solutions~\cite{pell_eq}. These can be generated recursively from the fundamental solution $(k_1, l_1) = (1, 1)$ via $k_{n+1} = 3k_n + 4l_n$ and $l_{n+1} = 2k_n + 3l_n$, for $n \geq 1$.
    
    The first four positive solutions are $(7, 5)$ with $\overline{z} = 24$, $(41, 29)$ with $\overline{z} = 840$, $(239, 169)$ with $\overline{z} = 28560$, and $(1393, 985)$ with $\overline{z} = 970224$. For $\overline{z} = 24$, our result for the diamond-shaped system with $r = 3$ ($k = 7$) yields $p_c = 0.16132(1)$, while the $5 \times 5$ ($l=5$) aligned square gives $p_c = 0.1647124(6)$~\cite{xun}, approximately $2\%$ higher.
    
    At higher values of $\overline{z}$, no numerical estimates of $p_c$ are available for either diamonds or aligned squares. However, the analytical results obtained by Jasna and Sasidevan~\cite{jasna2024} using the discrete excluded volume theory allow for estimating $p_c$ for both configurations at arbitrary values of $\overline{z}$. These estimates are expected to become increasingly accurate as $\overline{z}$ grows. Therefore, for $\overline{z} = 840$, $28560$, and $970224$, the predicted values from their approach should be very close to the true percolation thresholds. 

    By substituting the values of $\overline{z}$ for the diamond and aligned square configurations into Eqs.~\eqref{jasna_diamond} and~\eqref{jasna_square}, respectively, we obtain the following results: for $\overline{z} = 840$, the estimated thresholds are $p_c = 0.00511104$ (diamonds) and $p_c = 0.00506399$ (squares), with the diamond value being approximately $0.9\%$ higher; for $\overline{z} = 28560$, $p_c = 1.532610 \cdot 10^{-4}$ (diamonds) and $p_c=1.529996 \cdot 10^{-4}$ (squares), a difference of about $0.2\%$; and for $\overline{z} = 970224$, $p_c = 4.527017 \cdot 10^{-6}$ (diamonds) and $p_c=4.525674 \cdot 10^{-6}$ (squares), a relative difference of only $0.03\%$. Since the continuum percolation threshold $\eta_c$ is the same for both aligned squares and diamond-shaped objects on the square lattice, and since $\overline{z}$ exhibits similar asymptotic behavior in both configurations, the difference between the corresponding values of $p_c$ is expected to vanish as $\overline{z} \to \infty$.

    \subsection{Neighborhood of sizes $0$ and $1$}

\begin{figure}[h!]
        \centering
        \includegraphics[width=14.0cm, height=9.5cm]{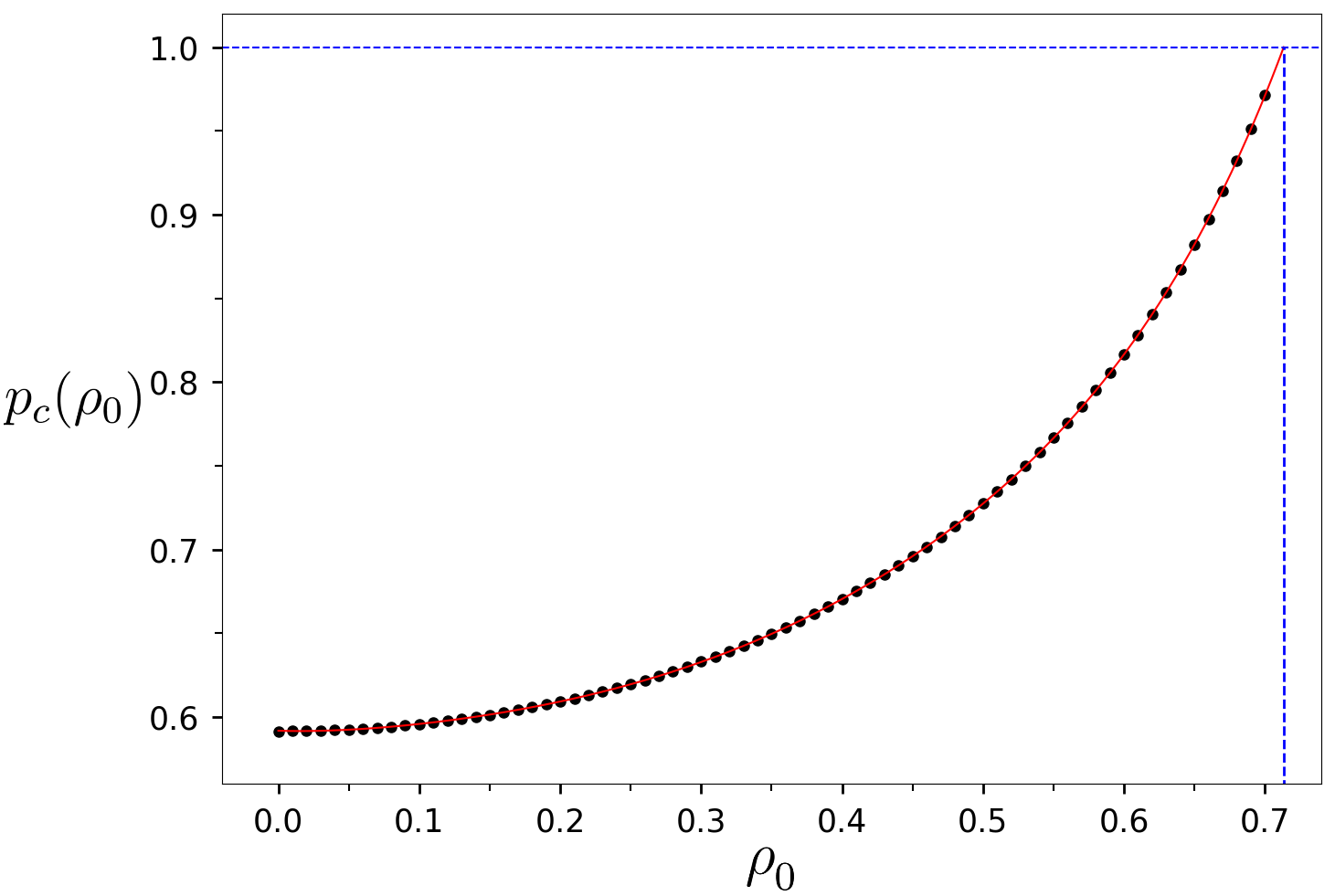}
        \caption{The critical percolation threshold $p_c$ is shown as a function of the fraction $\rho_0$ of sites with neighborhood radius 0. The system is restricted to neighborhood radius of 0 and 1 on a lattice of size $L=1024$. The points represent simulation data, showing a monotonic increase of $p_c$ with $\rho_0$. The point where the horizontal line ($p_c(\rho_0) = 1$) intersects the fitted curve gives $\rho_0^{c} \approx 0.715$, beyond which percolation becomes impossible. The error bars are smaller than the symbols.}
        \label{fig_rho_0}
    \end{figure}

    We consider that the system is populated by a fraction $\rho_0$ of sites with neighborhood size 0 and a complementary fraction $\rho_1 = 1 - \rho_0$ of sites with neighborhood size 1.  We vary $\rho_0$ and measure its effect on the critical percolation threshold, $p_c(\rho_0)$. The results of this analysis are presented in Fig.~\ref{fig_rho_0}, based on simulations with $10^5$ samples, system size $L=1024$, and increment $\Delta \rho_0 = 0.01$.
    
    As expected, the plot shows that $p_c(\rho_0)$ increases monotonically with $\rho_0$, indicating that a higher density of open sites is required to achieve percolation as the fraction of inert sites grows. By fitting the data with a rational function of the form  $y(x) = (a + bx)/(1 + cx + dx^2)$ we extrapolate the curve and estimate that $p_c(\rho_0) = 1$ when $\rho_0 = 0.715(1)$. Since the simulations were performed for a single system size ($L = 1024$), we interpret this value as an approximate estimate of the critical fraction $\rho_0^{c}$ such that, for any $\rho_0 > \rho_0^c$, percolation becomes impossible.

    This result shows that the sites with neighborhood radius $0$ actively facilitate percolation. If these sites did not contribute to forming clusters at all, percolation would be possible only if the density of the sites with neighborhood radius $1$, $\rho_1$, exceeded the critical threshold for the standard site percolation model on a square lattice, which is $p_c^{\text{site}} \approx 0.592746$. This implies that percolation would cease to be possible if $\rho_1 < 0.592746$, which corresponds to $\rho_0 > 0.407254$. However, our results demonstrate that the system percolates for values of $\rho_0$ well beyond this limit, $\rho_0^c \approx 0.715$. This shows that the ``0'' sites act as crucial ``bridges'' that connect sites with neighborhood radius 1, thereby aiding the formation of the spanning cluster and significantly extending the range over which percolation can occur.

    In fact, the value of $\rho_0^c$ can be related to a known percolation result by the following argument.  The bridges of the ``0'' sites mentioned above have the effect of increasing the range of interaction between ``1" sites to the first, second, and third nearest neighbors.  In this case, it is known that the critical concentration of the ``1" sites is $\rho_1^c = p_c(\textsc{sq-}1,2,3) = 0.2891226(14)$ \cite{xun}, which implies $\rho_0^c = 1 - \rho_1^c = 0.7108774(14)$. Our estimate, $\rho_0^c \approx 0.715(1)$, is slightly higher, which is expected given that it was obtained from simulations at a single system size ($L = 1024$), without finite-size scaling.

    \section{Relation to Continuum Percolation}

    For systems of a single diamond size, there is a direct connection between the extended-range percolation model with diamond-shaped neighborhoods, and the continuum percolation of aligned diamonds or equivalently squares.  Consider a percolation system on a lattice where we fill in lattice squares within the shape of a given object.  Here we consider the object to be a diamond of radius $r$ containing  $k = 2r + 1$ sites along the diagonal. This diamond-shaped object contains a total of 
    \begin{equation}
        n_r = 2r^2 + 2r + 1 = z_r + 1
    \end{equation}
    sites.  If we assign one index site to the center of the diamond, and $p$ the probability that that index site is occupied, then the net surface coverage $\phi$ is related to $p$ by
    \begin{equation}
        \phi(p) = 1 - (1-p)^{n_r} 
        \label{eq_phip}
    \end{equation}
    because for a given site to be unoccupied, neither it nor any  of the $n_r-1$ sites surrounding it can be occupied by the center of a diamond.  The above equation shows how the overall coverage of sites by the overlapping objects is related to $p$.
    
\begin{figure}[h!]
        \centering
        \includegraphics[width=12.5cm, height=12cm]{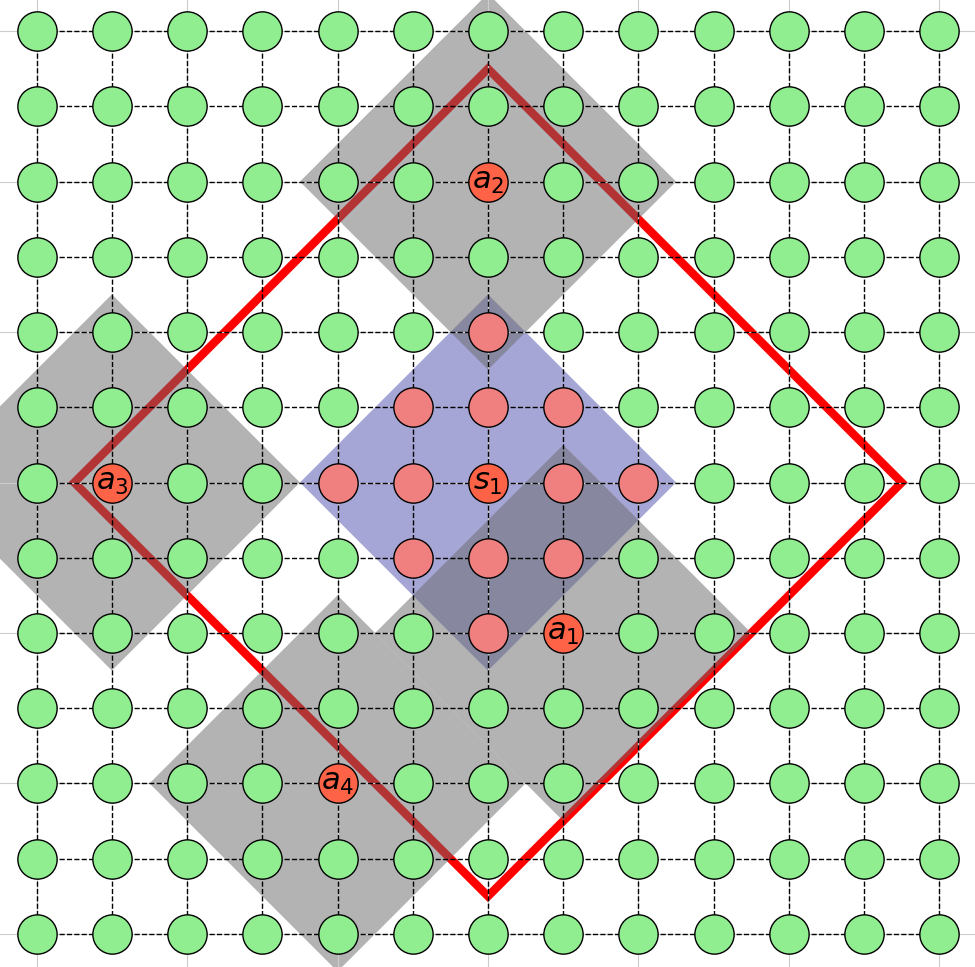}
        \caption{A region of a square lattice showing the central diamond of radius $r=2$, centered at site $s_1$. The diamond is indicated by the blue-shaded region composed of the red sites. The thick red line delineates a larger diamond-shaped neighborhood with an effective radius of $\tilde{r} = 2r+1 = 5$. Any other diamond of the same size ($r=2$) will touch or overlap the central one if its center is located within this red-bordered neighborhood. For instance, the gray diamonds centered at $a_1$ and $a_2$ overlap the central object, while the diamond centered at $a_3$ touches it at the boundary. Conversely, the diamond centered at $a_4$ has its center outside this region and therefore does not interact. This illustrates that two diamond-shaped objects of radius $r$ will touch or overlap if the distance between their centers is less than or equal to $2r+1$.}
        \label{fig_2r_plus_1}
    \end{figure}

    Next we look at the relation to extended-range percolation.  As shown in Fig.\ \ref{fig_2r_plus_1}, two diamond-shaped objects of radius $r$ will touch or overlap with each other if their centers are within a distance of $2r+1$ of each other, also in a diamond-shaped neighborhood of area $n_{2r+1}$, or in other words a neighborhood with coordination number of 
    \begin{equation}
        z_{2r+1} = 8r^2+12r+4.
    \end{equation}
    Here we find an interesting difference between this system and that of aligned squares of edge length $k$, where the area of a square object is $k^2$: in that case, the equivalent extended-range neighborhood is {\it not} of the same square shape, but a square with the corners cut off as seen in Ref.\ \cite{koza}, with an area of $(2k+1)^2-4$.  In the case of a diamond, the percolation neighborhood is identical to the shape of the continuum percolation objects, but larger.

    For example, based on Fig.~\ref{fig_2r_plus_1}, we can conclude that our model with a fixed neighborhood of radius $r=5$ (i.e., $i = m = 5$) is equivalent to the percolation of diamond-shaped objects with radius $r = 2$ on a lattice.

    Similarly, for $i = m = 9$, where $p_c = 0.02398(1)$ and $z = 180$ (Table~\ref{table_pc_i_i}), the neighborhood corresponds to the percolation of diamond-shaped objects with radius $r = 4$.
    For this system, we have for the net coverage of sites,
    \begin{equation}
        \phi = 1 - (1-p_{c,9})^{n_4} = 0.63034
    \end{equation}
    where $p_{c,9} = 0.02398(1)$ and $n_4 = z_4 + 1 = 41$.  
       
    For a discrete system one defines $\eta$ as the total area of all the objects, overlapping or not:
    \begin{equation}
        \eta = n_r p_c = (41)(0.02398) = 0.98318
        \label{eq_eta}
    \end{equation}
    for $r = 4$.
    This result compares to the approximation of Eq.\ (\ref{cont})
    \begin{equation}
        \eta_c = zp_c/4 = (180)(0.02397)/4 = 1.0791
        \label{eq_eta_c}
    \end{equation}
    which interestingly provides a better estimate of the continuum value for aligned squares of $\eta = 1.098843$.  Note that Eq.\ (\ref{eq_eta_c}) follows from Eq.\ (\ref{eq_eta}) by virtue of the fact that $n_r \approx z/4$ where $r$ represents the size of the object and $z$ represents the coordination number of the effective neighborhood.
    
    We note that combining Eqs.\ (\ref{eq_eta}) and (\ref{eq_phip}) yields 
    \begin{equation}
            \phi = 1 - (1-\eta /n_r)^{n_r} \sim 1-e^{-\eta}
            \label{eq_phieta}
        \end{equation}
    as $r \to \infty$, which represents the continuum limit here.
        
    Where there is a distribution of sizes, of diamonds on a surface, the effective neighborhood around a given deposited object depends not only on the size of the neighboring object.  Thus, it seems that the distribution of neighborhood sizes considered here (as indicated by $i, m$) does not directly relate to the continuum percolation of diamonds of different sizes.
    
    We note that for the continuum percolation of disks, studies have been made of systems of a uniform distribution of radii up to an upper cutoff.  In that case, it was found that the coverage of the surface at the critical point, $\phi_c=0.686610(7)$ \cite{quintanilla_2001}, is slightly higher than the critical coverage for disks of uniform size, $\phi_c = 0.676348$ \cite{quintanilla_ziff_2007,mertens}.  This is in contrast to the case here, where having a distribution of diamond neighborhoods leads to a lower critical point (in terms of $\eta$, which is monotonic in $\phi$), highlighting the difference between having a distribution of neighborhood radii and having a distribution of overlapping object sizes.

    To make these points more clear, we have also carried out simulations directly of diamond-shaped groups of sites deposited on a lattice. We considered a mixture consisting of a fraction $1 - d_1$ of diamond objects with $r = 0$ (single site) and a fraction $d_1$ with $r = 1$ (five sites). For $r = 1$, the diamond object has the same shape as the neighborhood of site $s_2$ shown in Fig.\ \ref{diamonds}.

\begin{figure}[httb]
		\begin{center}\includegraphics[width = 5.5 in]{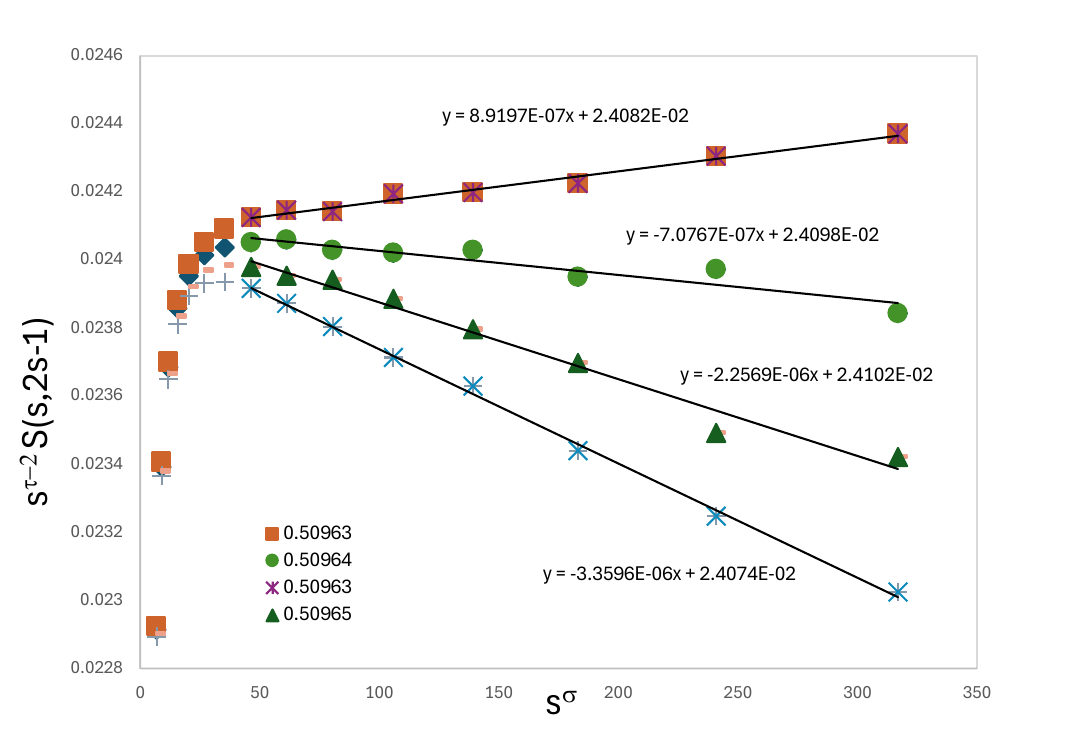} 
            \caption{Plot of $s^{\tau-2}S(s,2s-1)$ vs.\ $s^\sigma$ with $\tau = 187/91$ and $\sigma = 36/91$ for mixtures of diamond objects of $r=0$ and $r = 1$, with $d_1 = 0.1$ and values of $p$ shown in the legend.  The slopes are plotted in Fig.\ \ref{pcsloped01}.}
        \label{Sdo1}
    \end{center}
    \end{figure}

    With probability $p$, each site on the lattice was occupied by either a diamond with $r = 0$ (only the site itself was occupied), chosen with probability $1 - d_1$, or a diamond with $r = 1$ (the site and its four nearest neighbors were occupied), chosen with probability $d_1$. We then used a simple neighbor search algorithm to find the size distribution of the clusters in the system, assuming nearest neighboring occupied sites connect as in ordinary site percolation.  We recorded these results by the quantity

    \begin{equation}
        S(s,2s-1) = \sum_{s'=s}^{2s -1} s' n_{s'}(p)
    \end{equation}
    where $n_{s}(p)$ is the number of clusters of occupied sites of size $s$.  Note we consider $s'n_{s'}$ because this represents the total number of sites of occupied clusters of that size, and is found to give better statistics than just counting the number of clusters in that size range.  For large $s$ we expect 
    \begin{equation}
        S(s,2s-1) \sim s^{2-\tau} f(B(p-p_c)s^\sigma) \approx s^{2-\tau} (A + B (p-p_c)s^\sigma)
    \end{equation}
    where the latter applies for $p$ close to $p_c$. Thus we make plots of $s^{\tau-2} S(s,2s-1)$ vs.\ $s^\sigma$, and identify $p_c$ as the value of $p$ that makes the behavior horizontal for large $s$. When $p \ne p_c$, the behavior is linear with a slope proportional to $(p-p_c)$. In Fig.\ \ref{Sdo1}, we present this plot for the case $d_1 = 0.1$.
    
\begin{figure}[httb]
		\begin{center}\includegraphics[width = 5.5 in]{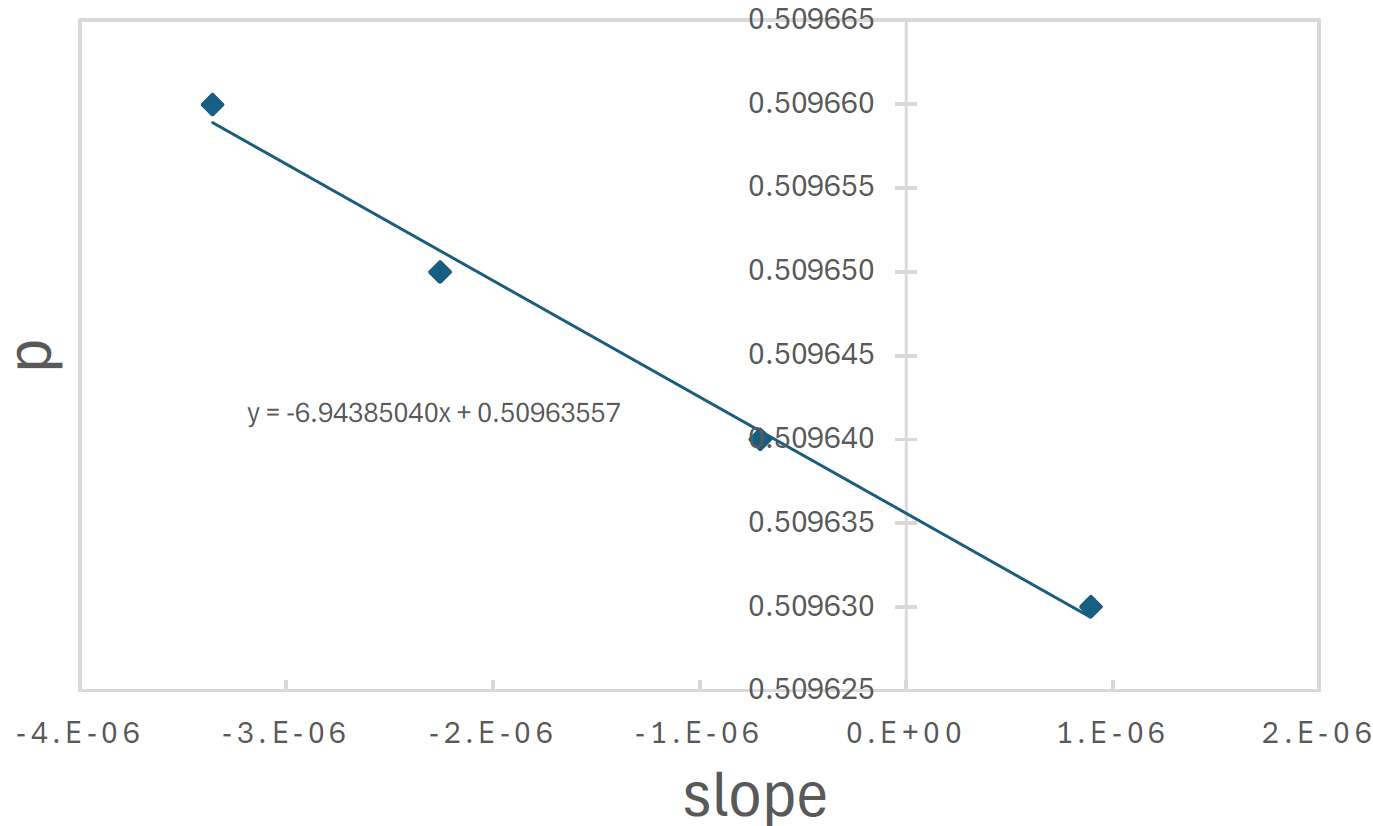} 
            \caption{Percolation probability $p$ vs.\ slopes of the asymptotic linear fits in Fig.\ \ref{Sdo1}.  The intercept for slope = 0 yields $p_c = 0.509636$ (for $d = 0.1$).} 
        \label{pcsloped01}
    \end{center}
    \end{figure}

    In Fig.\ \ref{pcsloped01} we see the slopes of the large-$s$ behavior of the plots.  The intercept where the slope is zero yields
    \begin{equation}
    p_c = 0.509636(1).
    \end{equation}
    \noindent The error bars, estimated from the fluctuation of the data, are consistent with the lower bound of the statistical error [lattice sites visited]$^{-1/2} = (1.26\cdot 10^5 \cdot 8192^2)^{-1/2} \approx 4 \cdot 10^{-7}$.

    We estimated $p_c$ considering $d_1 \in \{0, 0.1, 0.25, 0.5, 0.75, 1\}$, and the values obtained are shown in Table~\ref{table_d1}. For mixtures, $0<d_1<1$, the net coverage $\phi$ is related to $p_c$ by
    \begin{equation}
        \phi = 1 - (1 - p_c)(1 - d_1 p_c)^4.
        \label{eq_phid1}
    \end{equation}
    that is, one minus the probability that the site is empty and the four neighboring sites are also not occupied by an $r = 1$ diamond.
    For $d_1 = 1$, we obtained $p_c=0.161337(8)$ that is consistent with the value $p_c = 0.16132(1)$ found in our calculation above for the equivalent neighborhoods of size $r = 3$. Our result is also consistent with that of Malarz~\cite{malarz_2023}, $0.16134$. In this case, we have $\phi = 1 - (1 - p_c)^5 = 0.585106$, which is somewhat lower than the critical threshold of Bernoulli site percolation on the square lattice ($p_c = 0.592746$).

    \begin{table}[h!]
    \renewcommand{\arraystretch}{1.2} 
    \setlength{\extrarowheight}{0.25cm} 
    \setlength{\tabcolsep}{3.5pt} 
    \centering
    \begin{tabular}{|l|l|l|l|l|}
       \hline
        $d_1$ & $p_c$   & $\phi$ (Eq.\ \ref{eq_phid1}) & $\eta$ (Eq.\ \ref{eq_etad1})& $f$ (Eq.\ \ref{eq_vd1})\\
        \hline
        0  &  0.592746  & 0.592746 & 0.592746  & 1\\
        0.1  &  0.509636 & 0.602213 & 0.713489 & 0.642857\\
        0.25 & 0.396772	 &0.602801 & 0.793544 & 0.375\\
        0.5  & 0.272568  & 0.595168 & 0.817704 & 0.166667\\
        0.75  & 0.203365	 & 0.589068 & 0.813460 & 0.0625\\
        1.0  & 0.161337	 & 0.585106 & 0.806685 & 0 \\
        \hline
    \end{tabular}
    \caption{Quantities $p_c$, $\phi$, $\eta$, and $f$ for mixtures of components with fraction $d_1$ of $r = 1$-sized diamonds.}
    \label{table_d1}
    \end{table}

    For $d_1 = 0.5$, mixtures with an equal fraction of diamond objects of $r = 0$ and $r = 1$, we obtained $p_c = 0.272568(8)$. This does not agree with the results above, such as the $(1,2)$ result ($p_c = 0.33235(1)$) from Table~\ref{table_pc_i_m}, highlighting that this is indeed a different model. In this case, the net fraction of covered sites is $\phi = 0.595168$, which interestingly is just slightly above the threshold for $r = 0$ alone, where $\phi = p_c = 0.592746$.
   
    One can also look at $\eta$, the total area of all the deposited objects.  For the binary mixture, we have
    \begin{equation}
        \eta(d_1) = (1 + 4 d_1)p_c
        \label{eq_etad1}
    \end{equation}
    and the corresponding values, as well as the values of $\phi$, are also reported in Table~\ref{table_d1}. This function is concave with respect to $d_1$.

    In studies of continuum percolation of fully penetrable disks of two different radii, investigators have considered the behavior of the system as a function of the volumetric proportion of smaller disks as well as the overall density of disks.  In \cite{consiglio_2004} it was found that the behavior has a convex parabolic behavior, and in \cite{quintanilla_ziff_2007} it was shown that the behavior is not exactly symmetric.  
    
\begin{figure}[ht!]
        \centering
        \includegraphics[width=14cm, height=10cm]{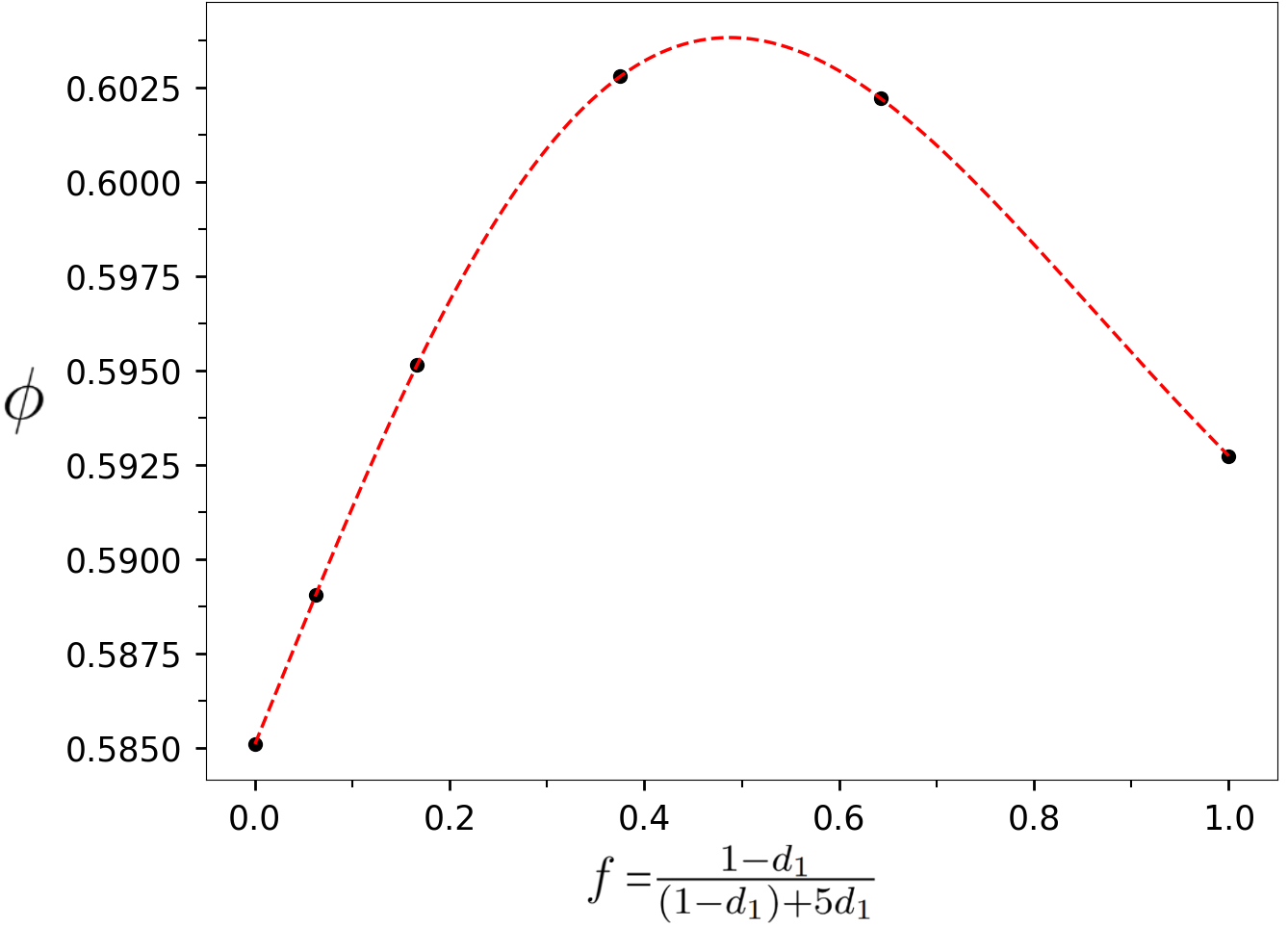}
        \caption{Plot of the percolation parameter $\phi$ as a function of the area proportion $f$ of smaller objects, where $(1-d_1)$ is the fraction of single-site objects. The black circles represent the simulated data points, and the red dashed line corresponds to a degree-5 polynomial fit given by $\phi(f) = -0.06333f^5+0.21798f^4 -0.22381f^3+0.01342f^2+0.06339f+0.58511$.}
        \label{phi_v}
    \end{figure}

    Following a similar approach, we investigated the behavior of $p_c$ and $\phi$ in our lattice model as a function of the area proportion of smaller objects, defined as \begin{equation}
        f(d_1) = \frac{1 - d_1}{(1 - d_1) + 5d_1} = \frac{1-d_1}{1+4d_1}
        \label{eq_vd1}
    \end{equation}
    where $d_1$ is the fraction of deposited objects that occupy five sites ($r=1$), and the remaining fraction $(1 - d_1)$ corresponds to objects occupying a single site ($r=0$). Figure \ref{phi_v} shows the plot of $\phi(f)$. Performing a degree-5 polynomial fit, we obtain $\phi(f) = -0.06333f^5+0.21798f^4 -0.22381f^3+0.01342f^2+0.06339f+0.58511$ (dashed curve). The resulting curve shows that $\phi(f)$ is asymmetric with respect to $f = 0.5$. Its maximum value is $\phi = 0.60383$, occurring at $f = 0.48770$.

\begin{figure}[h!]
    \begin{center} \includegraphics[width=0.8\textwidth]{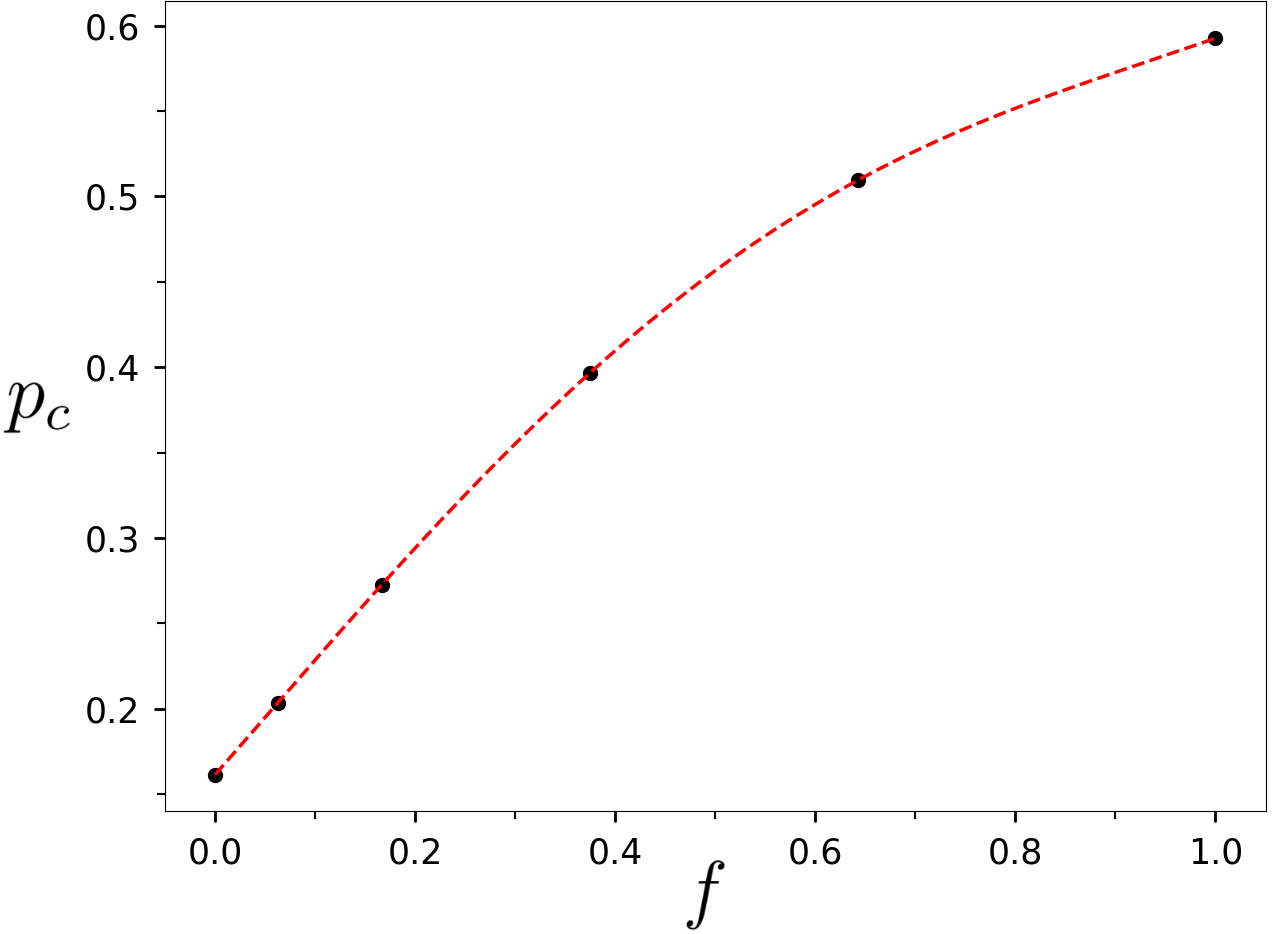} 
        \caption{Percolation threshold, $p_c$, as a function of the area proportion of smaller objects, $f$. The data points (black circles) are taken from Table~\ref{table_d1}. The red dashed curve is a fourth-degree polynomial fit to the data, given by the equation $p_c(f) = 0.35189f^4 - 0.69096f^3 + 0.10254f^2 + 0.66792f + 0.16135$.}
        \label{figpcf}
    \end{center}
    \end{figure}

    Figure~\ref{figpcf} illustrates the percolation threshold $p_c$ as a function of $f$. As expected, $p_c(f)$ is a monotonically increasing function, since a larger proportion of smaller objects hinders the formation of a percolating cluster. To model this behavior, we fitted the data with a fourth-degree polynomial, shown as the red dashed curve. The fit is described by the function $p_c(f) = 0.35189f^4 - 0.69096f^3 + 0.10254f^2 + 0.66792f + 0.16135$.
    
    We note that an alternative method to deposit particles or objects would be to allow them to overlap each other completely, including multiple adsorptions on a single site.  However, other than increasing the time of the simulation, this has no effect on the final surface coverage $\phi$ --- no correlations are introduced by this procedure. This fact is true not only for monodisperse objects, but also for distributions such as the $r=0$, $r=1$ mixture considered here.

    \section{Conclusion}

    We investigated a site percolation model on the square lattice in which each site is connected to others within a random compact diamond-shaped region of radius $r$, with $r$ drawn uniformly from the set $\{i, i+1, \ldots, m\}$, where $i = 0, \ldots, 7$ and $m = i, \ldots, 10$. For each pair $(i,m)$, we calculated the average number of neighbors per site, $\overline{z}(i,m)$, and estimated the critical percolation threshold $p_c(i,m)$ using finite-size scaling methods. When $i = m$, in addition to the values $i = 1, \ldots, 7$, we also considered $i = 8$, $9$, and $10$.

    In the case of identical neighborhoods ($i = m$), our results confirmed the expected convergence of $\overline{z}(i)\cdot p_c(i)$ toward $2^d \eta_c$ as $i \to \infty$, with $d = 2$ and $\eta_c$ denoting the critical threshold for continuum percolation of aligned square- (or diamond-) shaped objects.

    When considering random neighborhood sizes, we analyzed the behavior of the product $\overline{z}(i,m)\cdot p_c(i,m)$ for fixed $i$ and $m$ ranging from $i$ to $10$. In this case, we found numerical evidence that this product converges to a constant, denoted by $2^d \eta(i)$, as $m \to \infty$, indicating a behavior analogous to the identical neighborhood case. However, unlike the monodisperse case, where this limit matches the threshold of a corresponding continuum model of deposited objects, $\eta(i)$ does not appear to have a direct correspondence with typical continuum percolation models, although it may be related to a continuum version of the model defined here.
    
    Furthermore, our results highlight that not only the average number of neighbors but also the spread in neighborhood radius affects the critical behavior. Systems with similar average degree but broader distributions of neighborhood radius tend to percolate at lower occupation probabilities, as larger neighborhoods facilitate long-range connections.
    
    In the case of identical neighborhoods, we also compared percolation thresholds for diamond-shaped and aligned square configurations that share the same average degree $\overline{z}$. These configurations correspond to specific pairs of neighborhood parameters $(k, l)$ that satisfy Eq.~\eqref{eq_sq_diam}. Our results show that for $\overline{z} = 24$, the critical threshold for the diamond configuration ($r = 3$) is about $2\%$ lower than that of the aligned square ($5 \times 5$). However, starting from the next value of $\overline{z}$ ($\overline{z} = 840$), the analytical estimates based on the discrete excluded-volume theory in~\cite{jasna2024} indicate that $p_c$ remains slightly higher for diamonds.

    Finally, we also considered systems of deposited diamond-shaped objects, which for a monodisperse system is equivalent to the diamond-shaped neighborhood studied here, but for mixtures of diamonds of different sizes is not equivalent to the mixtures of neighborhoods studied here.  For example, for diamonds of size $r=0$ and size $r = 1$, we find the threshold is at $p_c = 0.272568(8)$, compared with $p_c = 0.592746$ for pure $r = 0$ and $p_c = 0.161337$ for pure $r=1$.  These correspond to a net coverage of $\phi = 0.595168$ compared with $\phi = 0.592746$ ($r = 0$) and $\phi = 0.585106$ for $r = 1$.  For mixtures of disks of different sizes it is known that mixtures lead to to a higher $\phi$ than disks of pure sizes, similar to what is seen here.  For the disks, it is known that the  distribution is close to parabolic \cite{consiglio_2004} but not exactly symmetric \cite{quintanilla_ziff_2007}.  An area for future study is the behavior for diamonds with different sizes and concentrations of the two species.

    \section*{Acknowledgements}
    Charles S. do Amaral was partially supported by FAPEMIG (Processo APQ-00272-24).

\end{document}